\documentclass[sigconf,balance=false]{acmart}
\renewcommand\footnotetextcopyrightpermission[1]{}

\usepackage{scalerel}
\usepackage{algorithmic}
\usepackage{caption}
\usepackage{subcaption}
\usepackage{bbm}
\usepackage{booktabs, multirow}
\usepackage{float}
\usepackage{cancel}
\DeclareMathOperator*{\argmax}{argmax}
\DeclareMathOperator*{\argmin}{argmin}

\newtheorem{theorem}{Theorem}[section]

\copyrightyear{YYYY}

\acmYear{YYYY}
\acmVolume{YYYY}
\acmNumber{X}
\acmDOI{XXXXXXX.XXXXXXX}
\acmISBN{}

\begin{document}

\title{A Cautionary Tale:\\ On the Role of Reference Data in Empirical Privacy Defenses}


\author{Caelin G. Kaplan}
\orcid{1234-5678-9012}
\affiliation{%
  \institution{SAP Labs France, Inria, Université Côte d’Azur}
  \city{}
  \country{}}
\email{caelin.kaplan@sap.com}

\author{Chuan Xu}
\affiliation{%
  \institution{Univ. Côte d’Azur, Inria, CNRS, I3S}
    \city{}
  \country{}}
\email{chuan.xu@inria.fr}

\author{Othmane Marfoq}
\affiliation{%
  \institution{Inria, Université Côte d’Azur, Accenture Labs}
    \city{}
  \country{}}
\email{othmane.marfoq@inria.fr}

\author{Giovanni Neglia}
\affiliation{%
  \institution{Inria, Univ. Côte d’Azur}
    \city{}
  \country{}}
\email{giovanni.neglia@inria.fr}

\author{Anderson Santana de Oliveira}
\affiliation{%
 \institution{SAP Labs France}
   \city{}
  \country{}}
\email{anderson.santana.de.oliveira@sap.com}


\renewcommand{\shortauthors}{Kaplan et al.}

\begin{abstract}
    Within the realm of privacy-preserving machine learning, empirical privacy defenses have been proposed as a solution to achieve satisfactory levels of training data privacy without a significant drop in model utility. Most existing defenses against membership inference attacks assume access to reference data, defined as an additional dataset coming from the same (or a similar) underlying distribution as training data. Despite the common use of reference data, \textcolor{black}{previous works are notably reticent} about defining and evaluating reference data privacy. As gains in model utility and/or training data privacy may come at the expense of reference data privacy, it is essential that all three aspects are duly considered. In this paper, we conduct the first comprehensive analysis of empirical privacy defenses. First, we examine the availability of reference data and its privacy treatment in previous works and demonstrate its necessity for fairly comparing defenses. Second, we propose a baseline defense that enables the utility-privacy tradeoff with respect to both training and reference data to be easily understood. Our method is formulated as an empirical risk minimization with a constraint on the generalization error, which, in practice, can be evaluated as a weighted empirical risk minimization (WERM) over the training and reference datasets. 
    Although we conceived of WERM as a simple baseline, our experiments show that, surprisingly, 
    it outperforms the most well-studied and current state-of-the-art empirical privacy defenses using reference data
    \textcolor{black}{for nearly all  relative privacy levels of reference and training data}. Our investigation also reveals that these existing methods are unable to trade off reference data privacy for model utility and/or training data privacy, and thus fail to operate outside of the high reference data privacy case. Overall, our work highlights the need for a proper evaluation of the triad ``model utility / training data privacy / reference data privacy'' when comparing privacy defenses.
\end{abstract}


\maketitle

\fancyhead{}
\thispagestyle{fancy}
\renewcommand{\thefootnote}{*}
\footnotetext{This paper has been accepted to PETS 2024: The 24th Privacy Enhancing Technologies Symposium, July 15–20, 2024, Bristol, UK.}
\renewcommand{\thefootnote}{\arabic{footnote}}
\setcounter{footnote}{0} 

\section{Introduction}
Data-driven applications, often using machine learning models, are proliferating throughout industry and society. Consequently, concerns about the use of data relating to individual persons has led to to a growing body of legislation, most notably the European Union's General Data Protection Regulation (GDPR)~\cite{gdpr2018eu}. 
According to the GDPR principle of data minimization, it is necessary to reduce the degree to which data can be connected to individuals, even when that data is used for the purposes of training a statistical model~\cite{gdpr-ai2020eu}. 
It has therefore become important to ensure that 
a machine learning model is not leaking private information about its training data. 

Membership inference attacks (MIAs), which seek to discern whether or not a given data point has been used during training, have emerged as \textcolor{black}{a key evaluation tool} for empirically measuring a machine learning model's privacy leakage~\cite{shokri2017membership}. 
Indeed, inferring training dataset membership can be thought of as the most fundamental privacy violation. 
Although other attacks exist, such as model inversion~\cite{fredrikson2015model}, property inference~\cite{ganju2018property}, dataset reconstruction~\cite{salem2020updates}, and model extraction~\cite{he2021stealing,krishna2019thieves,tramer2016stealing}, they all require a stronger adversary than is necessary for MIAs. 

Many methods have been proposed to defend against MIAs. The use of differential privacy~\cite{dwork2014algorithmic} (DP) has emerged as a leading candidate for two reasons. First, it provides mathematically rigorous guarantees that upper-bound the influence a given data point can exert on the final machine learning model.
Second, it is straightforward to integrate DP into a machine learning model's training procedure with algorithms such as differentially private gradient descent (DP-SGD)~\cite{abadi2016deep} or PATE~\cite{papernot2016semi}. Despite the many advantages associated with DP, there are several key drawbacks that include: the significant degradation of model utility when using DP during training~\cite{tramer2020differentially}, even more severe for underrepresented groups~\cite{bagdasaryan2019differential,uniyal2021dp,ganev2022robin, de2023empirical}, and the difficulty of translating DP's theoretical privacy guarantees to real-world privacy leakage~\cite{carlini2022membership,bernau2019assessing,nasr2021adversary,ye2021enhanced}.

To address these issues, empirical privacy defenses (i.e., without theoretical privacy guarantees) have been developed to protect the privacy of training data against MIAs. Existing empirical privacy defenses can be categorized by their method of protecting the training data (e.g., regularization~\cite{li2021membership,nasr2018machine}, confidence-vector masking~\cite{jia2019memguard,yang2020defending}, knowledge distillation~\cite{tang2021mitigating}). Alternatively, one can group defenses by whether they use only the private training data~\cite{tang2021mitigating} or require access to reference data~\cite{nasr2018machine,li2021membership,shejwalkar2021membership,jia2019memguard, yang2020defending,wang2020against}, defined as additional data from the same (or a similar) underlying distribution~\cite{nasr2018machine}. The two most prominent differentially private defenses can also be distinguished according to this distinction, where PATE~\cite{papernot2016semi} requires access to (unlabeled) reference data but DP-SGD~\cite{abadi2016deep} does not. 

\begin{figure}
\includegraphics[scale=0.44]{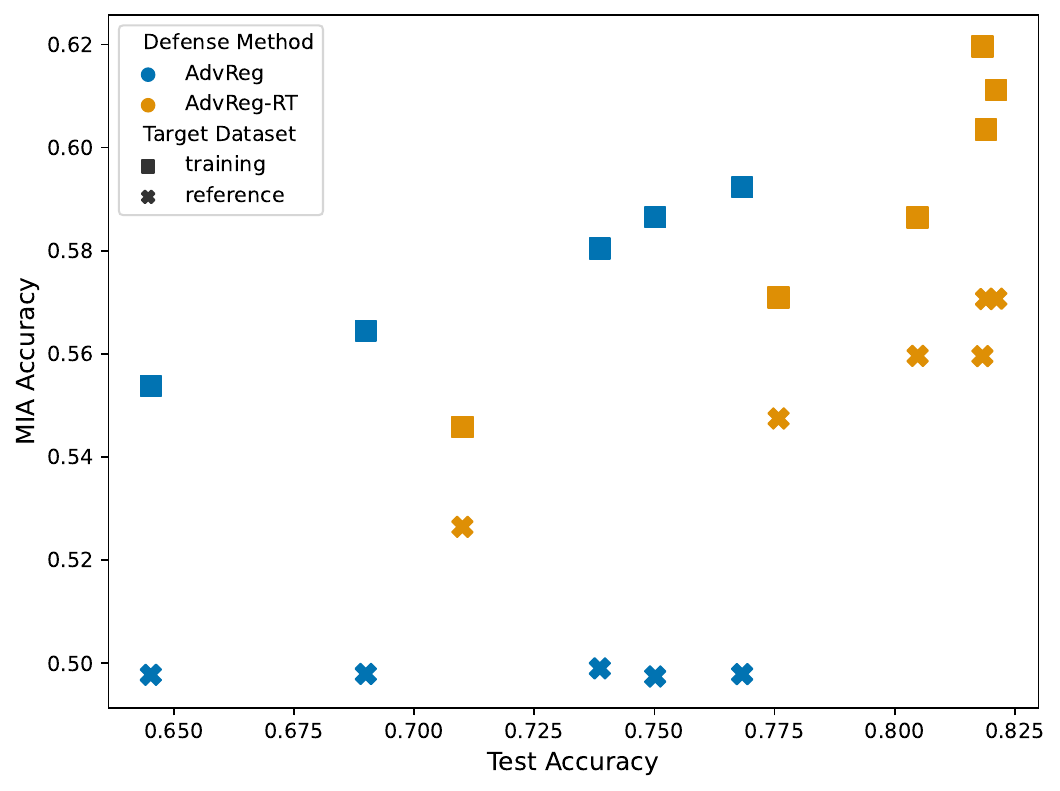}
\caption{Tradeoff between a defended classifier's prediction accuracy on test data (i.e., its model utility), MIA accuracy on training data (i.e., training data privacy leakage), MIA accuracy on reference data (i.e., reference data privacy leakage) for Purchase100 dataset. The key takeaway is that one cannot solely look at training data privacy leakage when evaluating the utility-privacy tradeoff of a given defense method.}\label{fig:adv_reg-tradeoff}
\Description{The utility-privacy curves for AdvReg and AdvReg-RT are compared in order to show that one cannot determine the better method without knowing the relative privacy considerations between training and reference data.}
\end{figure}

There are several problems with the current evaluation strategy of empirical privacy defenses. First, today's best practice is to produce a utility-privacy curve that compares a model's classification accuracy with its training data privacy for different values of a given defense parameter (e.g., different regularization term values).
Although this approach appears valid in the general case, assuming access to reference data makes the situation more complicated. \textcolor{black}{This additional dataset may have its own privacy requirements~\cite{li2021membership, shejwalkar2021membership,yang2020defending}, which we discuss in detail in Section~\ref{sec:ref-data-treatment}.} As gains in model utility and/or training data privacy 
usually come at the expense of reference data privacy, it is only possible to meaningfully compare defenses when the \textit{relative} level of privacy considerations between these two datasets is made explicit. \textcolor{black}{To demonstrate this issue, we present a concrete example in Figure~\ref{fig:adv_reg-tradeoff}, where ``AdvReg'' corresponds to adversarial regularization~\cite{nasr2018machine}, the most well-studied empirical privacy defense, and ``AdvReg-RT'' corresponds to an alternative version of the defense that we propose (defined in Section~\ref{sec:adv-reg-exp}).}
Looking only at the utility-privacy curves\footnote{
        For the AdvReg and AdvReg-RT, the curves are obtained by changing the relative importance of the classification loss and the attacker loss~\cite{nasr2018machine}, \textcolor{black}{i.e., the value of the parameter $\lambda$ in \eqref{adv-reg-opt}}.
} with respect to training data, 
it seems that AdvReg-RT is strictly better than AdvReg: for any given value value of test accuracy, AdvReg-RT is able to produce a model that yields a lower MIA accuracy on the training data.
However, when the utility-privacy curves are examined with respect to both training and reference data, one cannot determine the better method without knowing their relative privacy considerations. 

A second problem with the current evaluation methodology is the lack of a well-understood and simple baseline. The literature contains several examples where proposed empirical privacy defenses have been later shown to leak significantly more training data privacy than originally reported and sometimes to even perform worse than simpler defenses~\cite{choquette2021label,li2021membership,song2021systematic}. 
A well-established baseline could have provided more accurate expectations about the ability of these defenses. 

\textit{Thus, there is a strong need for the development of a baseline designed to operate in the same assumption setting as the vast majority of existing empirical privacy defenses and for an evaluation that takes reference data privacy into account.}


\textbf{Contributions.} We introduce the notion of a training-reference data privacy tradeoff and conduct the first comprehensive investigation into how empirical privacy defenses perform with respect to all three relevant metrics: model utility, training data privacy leakage, and reference data privacy leakage. Given this evaluation setting, we propose a well-motivated baseline that introduces the privacy requirement as a constraint on the generalization capability~\cite{shalev2014understanding} of the learned model.  \textcolor{black}{
Our formulation leads to a convenient weighted empirical risk minimization (WERM), where the training and reference data can be weighted according to the relative privacy level of the two datasets.} We prove that WERM enjoys theoretical guarantees 
both on the resulting model utility and \textcolor{black}{the relative privacy level of training and reference data.}

Our experimental results show that, surprisingly, WERM outperforms state-of-the-art empirical privacy defenses using reference data in nearly all \textcolor{black}{training and reference data relative privacy regimes, including the case of public reference data}. Additionally, we demonstrate that existing methods are only capable of extracting limited information from reference data during training and thus fail to effectively trade off reference data privacy for model utility and/or training data privacy. In particular, 
the mechanisms provided by these defenses to control the utility-privacy tradeoff with respect to the three aforementioned factors do not function as expected, since they are only able to operate in the \textcolor{black}{case where reference data privacy is highly valued.} By contrast, WERM is interpretable, straightforward to train, and highly effective. These traits enable it to serve as a baseline for evaluating future empirical privacy defenses using reference data. 
Importantly, comparing against our method requires selecting relative weights for the loss on the training data and the reference data, which makes
explicit the underlying assumption about their relative privacy. 

The remainder of the paper is organized as follows. In Section~\ref{sec:background}, we provide the background knowledge necessary to understand the domain of empirical privacy defenses. In Section~\ref{sec:theory}, we present WERM and analyze its theoretical properties.  In Section~\ref{sec:experiments}, we conduct a comprehensive set of experiments to evaluate our baseline in comparison to existing state-of-the-art methods. 
In Section~\ref{sec:conclusion}, we conclude our paper and discuss future work.

\section{Background}\label{sec:background}
\subsection{Machine Learning Notation}
In standard classification tasks, the goal is to learn a function $f_{\theta}$ that maps a set of input examples $x \in \mathcal{X}$ to a k-class probability distribution over a set of classes $\mathcal{Y} = \{1, 2, \dots, k \}$. The function's output, $f_{\theta}(x)$, is a vector, known as the confidence-vector, where each entry, $f_{\theta}(x)_{y}$, represents the model's confidence 
about input $x$ belonging to class~$y$.
The model training entity has access to a training dataset of $n$ examples, $D_{T} = \{(x_{1}, y_{1}), \dots, (x_{N_{T}}, y_{N_{T}})\}$, which have been drawn from an unknown underlying distribution~$\mathbb{D}$.

Although we only have access to $D_{T}$, for a learned function to make useful predictions, it must perform well on unseen data also coming from $\mathbb{D}$ (i.e., test data).  More formally, the task of training a model entails finding the vector of parameters, $\theta \in \Theta$, that minimize the expected risk (expected loss)~$L_{\mathbb{D}}$:
\begin{equation}\label{exp-risk}
    \min_{\theta \in \Theta} \: L_{\mathbb{D}}(f_{\theta}) = \min_{\theta \in \Theta} \: \underset{(x,y)\sim \mathbb D}{\mathbb{E}} \: \left[\ell\!\left(f_{\theta},(x,y)\right) \right] 
\end{equation}
where $\ell(\cdot, \cdot)$ is a loss function. In supervised classification tasks, the loss function is often chosen to be the cross-entropy loss,
$\ell\!\left(f_{\theta}, (x,y)\right) \allowbreak= - \sum_{y' \in \mathcal Y} \mathbbm{1}_{y'= y}\log(f_{\theta}(x)_{y})$. As we do not have access to $\mathbb{D}$, we cannot directly minimize the expected risk. Therefore, we instead minimize the loss over our training data, $D_{T}$, which we define as the empirical risk (training loss)~$L_{D_{T}}$:
\begin{equation}\label{emp-risk}
    \min_{\theta \in \Theta}  \: L_{D_{T}}(f_{\theta}) = \min_{\theta \in \Theta}  \:\frac{1}{N_{T}}\sum_{i=1}^{N_{T}} \ell\!\left(f_{\theta}, (x_{i}, y_{i})\right). 
\end{equation}

The empirical risk minimization (ERM) in~\eqref{emp-risk} 
is often solved using gradient descent methods~\cite{lecun1998gradient}. Given a satisfactory set of model parameters, $\theta_{s}$, the generalization error, also referred to as the generalization gap, is defined as:
\begin{equation}\label{gen-error}
    \text{Generalization Error} = L_{\mathbb{D}}(f_{\theta_{s}}) - L_{D_{T}}(f_{\theta_{s}}).
\end{equation}
The generalization error serves to quantify the difference between the training loss and expected loss. The framework of statistical learning theory~\cite{shalev2014understanding} enables the derivation of 
theoretical bounds for the generalization error, \textcolor{black}{which we use to provide guarantees for our proposed method.}

\subsection{Membership Inference Attacks}
\subsubsection{Attack Setting}
In the most generic case, a MIA operates in a setting where there exists:
\begin{itemize}
    \item A training dataset, $D_{T}$, (drawn from the distribution $\mathbb D$), whose privacy should be protected
    \item A machine learning model, $f_{\theta}$, which will be referred to as the target model, that is trained on $D_{T}$ and possibly additional data sources (e.g., reference data)
    \item An adversary, $\mathcal{A}$, who seeks to infer whether a target data point in a set $D^{\text{adv}}$ belongs to $D_{T}$
\end{itemize}

\subsubsection{Evaluation Setting}
The dataset $D^{\text{adv}}$, used for evaluating the performance of most previous attacks~\cite{nasr2018machine,shokri2017membership,song2021systematic,yeom2020overfitting}, is constructed such that it contains half of the training data, denoted as $D_{T}^{\text{adv}}$, and an equal size sample of non-training data from the same 
underlying distribution, denoted as $D_{\overline{T}}^{\text{adv}}$. \textcolor{black}{Accuracy is the standard metric used for evaluation, although recent work by Carlini et al.~\cite{carlini2022membership} proposes an alternative.}

We use the notation $\mathcal{A}(x,y)$ to define the binary output of a generic MIA, which codes members as 1 and non-members as 0. The accuracy of an attack against $D^{\text{adv}}$ can thus be calculated as:
\begin{equation}\label{mi-attack-acc}
    \frac{\sum_{(x_{i},y_{i}) \in D_{T}^{\text{adv}}}\mathcal{A}(x_{i},y_{i}) + \sum_{(x_{i},y_{i}) \in D_{\overline{T}}^{\text{adv}}}(1 - \mathcal{A}(x_{i},y_{i}))}{\bigl|D_{T}^{\text{adv}}\bigr| + \bigl|D_{\overline{T}}^{\text{adv}}\bigr|}
\end{equation}


\subsubsection{\textcolor{black}{Threat Model}}\label{sec:threat-model}
\textcolor{black}{
The potential for adversaries to perform effective membership inference increases with every additional piece of information they can access. Therefore, it is important to clearly articulate the assumptions underlying each potential attack. To the best of our knowledge, all known attacks proposed in the literature rely on at least one of the following four fundamental assumptions about the adversary's knowledge:
\begin{enumerate}
    \item Knowledge of the ground-truth label for a target data point.
    \item Access to either the largest confidence value or the entire confidence-vector when evaluated on a target data point, as opposed to merely the predicted label.
    \item Access to a dataset drawn from the same distribution as the training data (often referred to as population data~\cite{ye2021enhanced}).\footnote{\textcolor{black}{Note that the attacker's population data plays a similar role to reference data for the defender, but for clarity we avoid using the same name.}}
    \item Access to either a portion or all of the ground-truth training data, excluding the target data point whose membership the adversary wants to infer.
\end{enumerate}
Adversaries possessing access to either population data (Assumption~3) and/or ground-truth training data (Assumption~4) are positioned to launch significantly more sophisticated and potent attacks. In Table~\ref{table:attacks} (in Appendix~\ref{app:mia-assumptions}), we present the different adversary assumptions  for some of the most well-known MIAs.
}

\subsubsection{Existing Membership Inference Attacks}
MIAs can be levied against discriminative~\cite{shokri2017membership} and generative~\cite{chen2020gan} machine learning models. One key distinction among MIAs is whether the adversary has access to the inner-workings of the target model, such as weights, gradients, etc. (white-box), or only access to the target model's output (black-box). When evaluating our proposed baseline against state-of-the-art defenses, we follow previous work~\cite{jia2019memguard,nasr2018machine,tang2021mitigating} and assume that the adversary has black-box access to the target model. Therefore, from now on we focus on black-box attacks, and refer the reader to~\cite{nasr2019comprehensive} for a comprehensive review of white-box attacks. 

The simplest attack is the gap attack~\cite{yeom2018privacy}, which predicts any correctly classified data point as a member and any misclassified data point as a non-member:
\begin{equation}\label{gap-attack-single}
    \mathcal{A}_{\text{gap}}(x,y) = \mathbbm{1}\{\argmax_{i} f_{\theta}(x)_{i} = y\}.
\end{equation}
The name is derived from the fact that the attack directly exploits the generalization error (gap) described in \eqref{gen-error}. This attack only requires the assumption that an adversary has access to the ground-truth label. 

When the adversary has access to more fine-grained information (e.g., the confidence value associated to the predicted class or the entire confidence-vector), one can conduct a threshold-based attack~\cite{yeom2018privacy,song2021systematic}. Using the confidence value associated to the predicted class as an example, we have: 
\begin{equation}\label{thresh-attack}
    \mathcal{A}_{\text{conf}}(x,y) = \mathbbm{1}\{f_{\theta}(x)_{y} > \tau\},
\end{equation}
where $\tau$ is a class-independent threshold. \textcolor{black}{Song and Mittal~\cite{song2021systematic} demonstrated that threshold-based MIAs are the most effective 
among those that do not require access to training/non-training data. Further details regarding the design of the gap attack and extensions of threshold-based MIAs can be found in Appendix~\ref{app:mia-design}}. 

\textcolor{black}{In Section~\ref{sec:experiments}, following the methodology laid out in~\cite{song2021systematic}, we assess our proposed defense, Weighted Empirical Risk Minimization (WERM), against a variety of threshold-based MIAs. Additionally, we consider a neural network-based MIA~\cite{shokri2017membership}, which could be employed by a stronger adversary.}

\subsection{Empirical Privacy Defenses}
\label{sec:empirical_privacy_defenses}
Among empirical privacy defenses using reference data, the methods based on regularization techniques are the best performing~\cite{li2021membership}. As WERM belongs to this group, we provide background for this type of defense and refer the reader to~\cite{tang2021mitigating} for background on defenses using knowledge distillation. The idea of regularization defenses is to achieve a model that has good generalization, such that the distribution of model outputs on training data is similar to the output on unseen test data. Standard approaches to improve regularization, such as early-stopping~\cite{caruana2000overfitting}, weight decay~\cite{krogh1991simple}, and dropout~\cite{srivastava2014dropout}, have been observed to improve a model's robustness against a variety of MIAs~\cite{song2021systematic, shokri2017membership}. Additionally, some regularization terms have been proposed that seek to explicitly protect against attacks, such as adversarial regularization~\cite{nasr2018machine} and MMD-based regularization~\cite{li2021membership}. 

All empirical privacy defenses using reference data assume that the model training entity has access to training data, $D_{T} =\big\{(x_{1}, y_{1}),\allowbreak  \dots, (x_{N_{T}}, y_{N_{T}})\big\}$, and reference data, $D_{R} =\allowbreak \big\{(x_{1}^{\prime}, y_{1}^{\prime}), \dots, (x_{N_{R}}^{\prime}, y_{N_{R}}^{\prime})\big\}$, which come from the same (or a similar) underlying distribution and are of size $N_{T}=|D_{T}|$ and $N_{R}=|D_{R}|$, respectively. \textcolor{black}{The defenses aim to make model predictions on training and reference data sufficiently similar, such that it will be hard for an attacker to distinguish a model's output on training and non-training data. 
The closer the distributions of training data and reference data, the easier the task for the defense and the smaller the model utility loss.
}
\subsubsection{Adversarial Regularization}\label{sec:adv-reg-background}
Adversarial regularization (AdvReg)~\cite{nasr2018machine} is a model training framework that is formulated as a $\min \max$ game, where a classifier, $f_\theta$, is trained to be optimally protected against a MIA model, $h_\phi$. The first component is the loss of the classifier, $f_{\theta}$, over the training data, i.e.,~$L_{D_{T}}(f_{\theta})$ as described in~\eqref{emp-risk}.
The second component is the gain of the attack model: 
\begin{equation}\label{attack-component}
\begin{split}
    G_{D_{T}, D_{R}} (f_{\theta}, h_{\phi}) = \: &\frac{1}{N_{T}} \sum_{i=1}^{N_{T}} \log[h_{\phi}(x_{i}, y_{i}, f_{\theta}(x_{i}))] \: + \\
    &\frac{1}{N_{R}} \sum_{i=1}^{N_{R}} \log[1 - h_{\phi}(x_{i}^{\prime}, y_{i}^{\prime}, f_{\theta}(x_{i}^{\prime}))],
\end{split}
\end{equation}
where $h_{\phi}(x, y, f_{\theta}(x))$ outputs the probability that a given target data point is a member of the training data. The attack model's gain quantifies its ability to predict the training data as members and the reference data as non-members. 

The whole optimization problem can be formulated as:
\begin{equation}\label{adv-reg-opt}
    \min_{\theta \in \Theta}\max_{\phi \in \Phi} \: L_{D_{T}}(f_{\theta}) + \lambda \: G_{D_{T}, D_{R}}(f_{\theta}, h_{\phi}),
\end{equation}
where $\lambda$ is the regularization term's weight and serves to trade utility for privacy (a larger $\lambda$ should result in the trained model having greater privacy protection at the cost of decreased utility). The minmax problem described in~\eqref{adv-reg-opt} is solved by alternating some gradient method steps for the minimization and the maximization problem.

\subsubsection{MMD-based Regularization}\label{sec:mmd-background}
Alternatively, in MMD-based regularization (MMD) as proposed in~\cite{li2021membership}, the regularization term may be the Maximum Mean Discrepancy (MMD), leading to the following problem:
\begin{equation}
    \min_{\theta \in \Theta} L_{D_{T}}(f_{\theta}) +
     \lambda \cdot \Big\| \frac{1}{N_{T}} \sum_{i=1}^{N_{T}}\psi(f_{\theta}(x_{i})) - \frac{1}{N_{R}} \sum_{i=1}^{N_{R}}\psi(f_{\theta}(x^{\prime}_{i})) \Big\|_{\mathcal{H}}
    \label{eq:mmd}
\end{equation}
where $\mathcal{H}$ is a universal Reproducing Kernel Hilbert Space (RKHS) and $\psi$ is a function mapping model's outputs 
to points in $\mathcal{H}$.
By solving the problem in~\eqref{eq:mmd}, the resulting model  seeks to simultaneously minimize the empirical risk of the training data and the difference in output of the model on training and reference data in the space $\mathcal H$. 
Traditionally, to calculate the MMD one would find $\psi$ such that it maximizes the distance in $\mathcal{H}$. Instead, to simplify the training process, the authors of \cite{li2021membership} select~$\psi$ to be a given Gaussian kernel. 

\subsection{Reference Data Overview and Threat Model}\label{sec:ref-data-treatment}  
\textcolor{black}{The vast majority of empirical privacy defenses in the literature~~\cite{li2021membership,nasr2018machine,wang2020against,yang2020defending,jia2019memguard,papernot2016semi,papernot2018scalable} require access to reference data,
which is assumed to come from the same (or a similar) underlying distribution as 
training data. 
In Section~\ref{sec:ref-data-use-cases}, we discuss the availability of reference data and its level of privacy.
In Section~\ref{sec:ref-data-overview}, we examine how existing empirical privacy defenses have dealt with the privacy of reference data.
}
\subsubsection{Reference Data Availability and Privacy}
\label{sec:ref-data-use-cases}

\begin{table*}[hbt!]
\centering
\caption{Comparison of \textcolor{black}{empirical }privacy defenses by reference data treatment. In the third column, ``relative level unspecified'' means the target level of relative privacy requirements 
between training and reference data is not stated. In the fourth column, ``single privacy level'' means the reference data privacy leakage is evaluated at a single point on the utility-privacy curve. We use a dashed line (---) to convey that the defense either does not use reference data or does not need to evaluate reference data privacy leakage.}\label{table:ref-data}
\begin{tabular}{|p{5.1cm}|p{3.5cm}|p{4.4cm}|p{3.4cm}| }
\hline
Defense & Category & Reference Data Privacy Setting & Reference Data Privacy Evaluation \\
\hline
Adversarial Regularization~\cite{nasr2018machine} & regularization         & not mentioned & no evaluation\\
MemGuard~\cite{jia2019memguard} & confidence-vector masking                               & not mentioned & no evaluation\\
Model Pruning~\cite{wang2020against}  & knowledge distillation                     & not mentioned & no evaluation\\
MMD-based Regularization~\cite{li2021membership} & regularization         & private (relative level unspecified) & yes (single privacy level) \\
Distillation for Membership Privacy~\cite{shejwalkar2021membership} & knowledge distillation  & private (relative level unspecified) & yes (single privacy level)\\
Prediction Purification~\cite{yang2020defending}  & confidence-vector masking             & private (relative level unspecified) & yes (single privacy level)\\
\textcolor{black}{WERM (this paper)} & \textcolor{black}{regularization} & \textcolor{black}{all possible settings discussed} & \textcolor{black}{yes (all privacy levels)}\\
\hline
\end{tabular}
\end{table*}

Although not always called ``reference data,'' the notion of having access to a distinct dataset coming from the same (or a similar) underlying distribution as training data is common throughout many domains of machine learning literature \textcolor{black}{(e.g., the design of MIAs as mentioned in Section~\ref{sec:threat-model}). We can divide the examples into cases where reference data is public and cases where reference data is private.} In the public reference data setting, large publicly available datasets are routinely employed to pre-train a model which will later be fine-tuned using a private and smaller training dataset~\cite{bengio2012deep,choquette2021label} or a public dataset can be used for knowledge transfer across heterogeneous models trained on private local datasets in a federated learning scenario~\cite{lin2020ensemble, ijcai2021p205}.
\textcolor{black}{When reference data is public, empirical privacy defenses can use it to augment the privacy of training data, while disregarding concerns about the privacy of the reference data itself.} 

\textcolor{black}{
\textcolor{black}{In the private reference data setting, the availability of reference data may result from} model training entities having private datasets that contain certain records with distinct privacy requirements.  
The ``pay-for-privacy'' business model enables companies to acquire data from users or consumers at various privacy levels~\cite{elvy2017paying}. For example, ISPs are known to provide discounts to their users in exchange for the possibility of exploiting their data for targeted advertisement (possibly powered by a machine learning model)~\cite{federal2016protecting}, and some mobile phone applications offer a free and a paid version that provides better privacy protection to users of the paid service~\cite{han2020price}. Training and reference data can then correspond to data from users with a different pricing scheme. Different privacy levels may also be due to past data leaks, e.g., due to malicious security breaches or human errors. As will become apparent, in this scenario where a single dataset has two segments with distinct privacy considerations, one can use either the more or less private data segment as reference data to better protect the privacy of the remaining segment (training data). 
Even in standard machine learning training, such considerations may be a leading factor in choosing how to split the available data into training and validation segments, as they have been shown to each leak different amounts of private information~\cite{papernot2021hyperparameter}. Finally, we observe that  heterogeneity in privacy levels is also implicitly assumed in fog learning~\cite{hosseinalipour2020federated}, where federated learning clients share a part of their local datasets to bring their respective distributions closer to facilitate the training of a common model.
}

\subsubsection{Reference Data in Empirical Privacy Defenses}\label{sec:ref-data-overview}
 \textcolor{black}{In Table~\ref{table:ref-data}, we present seven empirical privacy defenses using reference data: the first six are existing defenses and the seventh is our proposed method, WERM. The existing defenses can be subdivided into two categories based on reference data privacy treatment: private~\cite{li2021membership,shejwalkar2021membership,yang2020defending} and ``not mentioned''~\cite{nasr2018machine,jia2019memguard,wang2020against}. 
 We use the label ``not mentioned'' to represent works where reference data privacy is neither discussed nor evaluated.}\footnote{
    \textcolor{black}{We note that the omission may suggest they implicitly consider the reference data to be public.}
} Moreover, each of the three works that consider reference data to be private evaluate its privacy leakage at only a single point on the utility-privacy curve and show it to be much smaller than 
the training data privacy leakage. 
These results 
reveal an implicit choice by the authors: reference data privacy is valued more highly than training data privacy.

We do not take a particular stance on the relative privacy of training and reference data, i.e., if the reference data in empirical privacy defenses should be considered more or less private than training data---as shown in Section~\ref{sec:experiments}, \textcolor{black}{we evaluate WERM in all possible reference data privacy settings} and show that it outperforms state-of-the-art defenses across almost the entire spectrum. 
Yet, we argue that, without quantifying the relative importance assigned to the three key objectives (model utility, training data privacy, and reference data privacy), we cannot adequately compare the performance of these defenses.
For example, in the papers that consider reference data more private than training data, the proposed defenses are still allowing for some reference data privacy leakage to achieve a high model utility and training data privacy protection. Is this the right amount of privacy leakage? Perhaps, one should instead seek to trade much more reference data privacy to improve the other two metrics. Alternatively, if reference data privacy is of the utmost importance, the current leakage may already be unacceptable. Similar considerations hold for the public reference data case: given that reference data privacy is not a concern, are the proposed methods achieving the best possible tradeoff between model utility and training data privacy?

\textcolor{black}{The next section will introduce our method and show how its utility-privacy tradeoffs are amenable to analysis.}
\vspace{-0.05cm}
\section{Weighted Empirical Risk Minimization}
\label{sec:theory}
In this section, we introduce 
our proposed baseline, WERM, and analyze its theoretical properties related to generalization and privacy protection. \textcolor{black}{WERM's design is rooted in the fundamental principles of statistical learning, particularly in the generalization error~\eqref{gen-error}. WERM utilizes a weight term, $w$, which simultaneously regulates the tradeoff between the privacy of training data and reference data, as well as the tradeoff between utility and privacy.} Employing tools from differential privacy (DP)~\cite{dwork2014algorithmic} and statistical learning theory~\cite{shalev2014understanding}, we derive theoretical bounds that enable us to understand how $w$ and the size of the two datasets impact the relative privacy leakage and the model's utility.
\textcolor{black}{Following all related work~\cite{nasr2018machine,li2021membership,wang2020against,jia2019memguard,shejwalkar2021membership,yang2020defending}, we consider the two distributions from which $D_{T}$ and $D_{R}$ are drawn to be identical. The relative privacy results in Theorem~\ref{th:privacy-theorem} do not depend on $D_{T}$ and $D_{R}$ coming from the same underlying distribution and the generalization bound in Theorem~\ref{th:gen-bound} can be extended to the case where the distributions are only similar.}
\subsection{Motivation}
\label{sec:motivation}
Drawing any conclusion about the quality of a defense can only come after comparing it to an interpretable and well-performing baseline. Therefore, our goal is to propose a baseline that makes the training-reference data privacy tradeoff explicit and can operate across the entire range of possible privacy settings. Our method's design originates from the understanding that all black-box MIAs share a common design feature, which is exploiting the difference between a model's output on training and non-training data. What they consider as a model's output may differ (e.g., predicted label, loss, confidence-vector), but the distinguishability of output distributions is the prerequisite for a membership inference vulnerability to exist in the black-box setting. Thus, employing an ideal membership inference defense will result in a defended model that behaves identically when queried with training or non-training data from the same distribution. The design of a defense based on regularization requires a decision about how to define equivalence of output. AdvReg (Section~\ref{sec:adv-reg-background}) introduces a regularization term that constrains the difference between a classifier's confidence-vector output on training and reference data based on a learned neural network; MMD (Section~\ref{sec:mmd-background}) constrains this difference using a Gaussian kernel. 
\textcolor{black}{Our proposed baseline is motivated by the fact that a smaller generalization error implies that the empirical loss is closer to the expected loss and, subsequently, the loss observed on any future sample drawn from the same distribution, making it difficult for the adversary to conclude which samples were part of the training data.} 
\textcolor{black}{Thus, 
WERM addresses the fundamental challenge common to all regularization defenses: learning a classifier whose outputs are indistinguishable between training and reference data.
    However, its design, rooted in statistical learning principles, results in a unique algorithm. WERM not only exhibits superior performance (Section~\ref{sec:empirical-results}) but also provides enhanced interpretability (Section~\ref{sec:method}), simpler configuration (Section~\ref{sec:weight-term-selection}), and reduced computational costs (Section~\ref{sec:training-time}).}


\subsection{Method}\label{sec:method}
We propose to train a standard ERM using both training and reference data, while constraining the generalization error with respect to each of the datasets. 
Our problem can be formulated as:
\begin{equation}\label{w-erm-formulation}
    \begin{aligned}
    \text{Input:} & \quad D_{T} \sim \mathbb{D}^{N_{T}}, D_{R} \sim \mathbb{D}^{N_{R}},  \; c_{T},c_{R}  \in \mathbb{R}^{+} \\
    \min_{\theta \in \Theta} & \quad L_{D}(f_{\theta}) \\
     \textrm{s.t.} &  \quad  L_{\mathbb{D}}(f_{\theta}) - L_{D_{T}}(f_{\theta}) \leq c_{T} \\
    \phantom{\textrm{s.t.}}  & \quad L_{\mathbb{D}}(f_{\theta}) - L_{D_{R}}(f_{\theta}) \leq c_{R}
    \end{aligned}
\end{equation}
where $D = D_{T} \cup D_{R}$, $N_{T}$ and $N_{R}$ are the respective sizes of training and reference data, $L_{D}(f_{\theta}) = \frac{N_T}{N} L_{D_{T}}(f_{\theta}) + \frac{N_R}{N} L_{D_{R}}(f_{\theta})$, with $N=N_T+N_R$, and the constants $c_{T}$ and $c_R$ constrain the generalization error on the training data and on the reference data, respectively. On the basis of our discussion in Section~\ref{sec:motivation}, smaller values of $c_T$ ($c_R$) correspond to greater privacy protection for training (reference) data.
For the purpose of readability, in the rest of this section, we write $L_{D}(f_{\theta})$ as $L_{D}$ (i.e., the loss over a given dataset is implied to be evaluated for $f_{\theta}$). Moreover, for simplicity, we consider the case where $N_T=N_R$.

\textcolor{black}{Studying the Lagrangian of problem~\ref{w-erm-formulation} and introducing the optimal multipliers $\lambda^{\ast}$ and $\mu^{\ast}$, as detailed in Appendix~\ref{app:lagrangrian}, we can show that~\eqref{w-erm-formulation} becomes equivalent to the following two problems:
\begin{align}
    \min_{\theta \in \Theta}& \: \Bigl[\frac{1}{2} + \mu^{\ast} \Bigr] L_{D_{T}} + \Bigl[\frac{1}{2} - \mu^{\ast} \Bigr] L_{D_{R}}\label{w-erm-train-priv}, \\
    \min_{\theta \in \Theta}& \: \Bigl[\frac{1}{2} - \lambda^{\ast} \Bigr] L_{D_{T}} + \Bigl[\frac{1}{2} + \lambda^{\ast} \Bigr] L_{D_{R}}\label{w-erm-ref-priv},
\end{align}
where~\eqref{w-erm-train-priv} corresponds to the case when reference data privacy is a stricter constraint (i.e., $c_{R} < c_{T}$) and~\eqref{w-erm-ref-priv} corresponds to the case when training data privacy is a stricter constant (i.e., $c_{T} < c_{R}$).}
In both cases, we obtain a weighted sum of the two empirical risks with a larger weight (i.e., $>1/2$) given to the dataset with looser privacy constraints. Using equal weights corresponds to equal privacy constraints.

Motivated by this reasoning, we propose the following weighted empirical risk minimization (WERM) as a baseline for privacy defenses using reference data:
\begin{equation}\label{w-erm-aggregated}
    \min_{\theta \in \Theta} \: L^{w}_{D}(f_{\theta}) = (1-w) L_{D_{T}}(f_{\theta}) + w L_{D_{R}}(f_{\theta}), \textrm{ for some }w \in [0,1].
\end{equation}
This formulation allows us to simply trade reference data privacy for training data privacy by changing the parameter $w$. Higher (lower) values of $w$ lead to greater privacy protection for training (reference) data. In particular, the privacy of training data and reference data is perfectly protected for $w=1$ and $w=0$, respectively, which is the case where the corresponding dataset is not used to compute the defended model. \textcolor{black}{
Another benefit of WERM's formulation in~\eqref{w-erm-aggregated} is its ability to accommodate multiple datasets, each with a distinct privacy level (up to the limit case where every point is a separate dataset with its own privacy considerations). It is unclear how AdvReg~\cite{nasr2018machine} or MMD~\cite{li2021membership} could be adapted to this scenario. We prove Theorem~\ref{th:privacy-theorem} on the relative privacy leakage (Appendix~\ref{app:proof-dp}) and Theorem~\ref{th:gen-bound} on the generalization bound (Appendix~\ref{app:proof-gen-bound}) for this generalized case.}

Along with its high interpretability, WERM is also a lightweight defense, as its computational cost is equivalent to training an undefended model by minimizing the empirical risk over $N$ samples. This is less computationally expensive than solving AdvReg's minmax problem in~\eqref{adv-reg-opt} and MMD's additional requirement of comparing the distance for unique classes in a batch (see implementation details in Appendix~\ref{app:mmd}). 
A detailed comparison of the training time for these defenses (Section~\ref{sec:training-time})
confirms this intuition.

In the remainder of this section, we provide theoretical guarantees for WERM's relative training-reference data privacy 
(Section~\ref{sec:privacy-analysis}) and WERM's
model utility (Section~\ref{sec:utility-analysis}).

\subsection{WERM's Privacy}\label{sec:privacy-analysis}
Our analysis in the previous section led us to qualitatively conclude that increasing (decreasing) the reference data weight, $w$, in WERM results in increased privacy protection for the training (reference) data. Particularly, when the two datasets are the same size and have the same privacy requirements, one should select $w=1/2$. 
In this section, we derive more formal privacy guarantees and configuration rules for $w$ considering general dataset sizes. 

\textcolor{black}{The formulation of WERM in~\eqref{w-erm-aggregated} is not intrinsically differentially private. However, using DP-SGD~\cite{abadi2016deep} as the optimization algorithm to solve~\eqref{w-erm-aggregated} enables WERM to become a differentially private method.
For the purpose of our analysis, we assume this situation in order to employ tools from DP~\cite{dwork2014algorithmic} to measure the relative privacy tradeoff between training data and reference data. 
Consequently, the $\epsilon$ values presented are simply a convenient way to achieve our primary goal of quantifying how the weight term, w, and the size of the two datasets impact WERM's relative privacy level. We emphasize that, while possible, we are not proposing to train WERM with DP-SGD to achieve $\epsilon$-DP privacy guarantees.
}


\textcolor{black}{DP-SGD works by clipping the gradient values below a certain threshold and adding Gaussian noise to each of them with scale $\sigma$. If properly configured, DP-SGD enjoys $(\epsilon, \delta)$-DP guarantees, i.e., 
when a single sample of the dataset is changed, the probability of any possible event observable by an attacker changes at most by a multiplicative factor $\exp(\epsilon)$ and by an additive term~$\delta$. The larger noise scale $\sigma$, the smaller $\epsilon\ge 0$ and $\delta 
\in [0,1)$, and the stronger the privacy guarantees.} 


\textcolor{black}{Fundamentally, an empirical privacy defense that has access to reference data must make a choice regarding how much of the reference data's privacy should be sacrificed to protect the privacy of the training data. We rely on the $\epsilon$ parameter from DP to quantify the relative privacy of the two datasets. As we will argue after stating our result, in practice, we can consider that the conclusions about the relative privacy hold even if DP-SGD is not used during training.
}

\begin{theorem}[Privacy Leakage]\label{th:privacy-theorem}
For some overall number of training steps, K, WERM minimized with DP-SGD is:
\begin{align}
    &\Bigl( O(\epsilon_{_{T}}), \delta \Bigr)-\text{DP w.r.t. the training dataset} \: (D_{T})\label{eq:train-priv} \\
    &\Bigl( O(\epsilon_{_{R}}), \delta \Bigr)-\text{DP w.r.t. the reference dataset} \: (D_{R})\label{eq:ref-priv}
\end{align}
where:
\begin{align}
    \epsilon_{_{T}} &= \epsilon_{_{0}} \frac{1-w}{N_{T}},
    \epsilon_{_{R}} = \epsilon_{_{0}} \frac{w}{N_{R}} \nonumber\\
    0 &< \epsilon_{_{0}} < \min\Bigl(\frac{N_{T}}{1-w}, \frac{N_{R}}{w}\Bigl), \label{eq:epsilon_zero}\\
    \sigma &= \alpha \sqrt{K} \sqrt{2 \log \frac{1.25}{\delta}} \frac{C}{\epsilon_{_{0}}}, \label{eq:sigma}
\end{align}
and $\sigma$, C, and $\alpha$ are the noise scale, gradient norm bound, and sampling ratio in DP-SGD, respectively. 
\end{theorem}

The proof of Theorem~\ref{th:privacy-theorem} and a detailed description of how we adapt the analysis of DP-SGD from Abadi et al.~\cite{abadi2016deep} to be compatible with WERM can be found in Appendix~\ref{app:proof-dp}. 

It is important to note that \textcolor{black}{the relative privacy of the two datasets, as quantified by the ratio $\epsilon_{_{T}}/\epsilon_{_{R}}$ 
is completely governed by $w$ and the size of the two datasets and independent of $\epsilon_0$}.  In particular, the training data will be more private if and only if $\frac{1-w}{N_{T}} < \frac{w}{N_{R}}$. Specifically, setting the weight of each empirical loss in~\eqref{w-erm-aggregated} proportional to the size of its corresponding dataset leads to the same privacy guarantees for samples in both datasets. In the case where $N_{T}=N_{R}$, we recover the result we were able to conclude qualitatively in the previous section, i.e., that setting $w=1/2$ will result in equivalent privacy guarantees for the training and reference data.

\textcolor{black}{The independence of the ratio $\epsilon_{_{T}}/\epsilon_{_{R}}$ on $\epsilon_0$ implies that the same value for the relative privacy of the two datasets is achieved if we set $\epsilon_0$ to a very large value (on the order of the dataset size, see~\eqref{eq:epsilon_zero}) and then use DP-SGD with a negligible noise ($\sigma \approx 0$ in~\eqref{eq:sigma}). These considerations justify our experimental results } 
in Section~\ref{sec:empirical-results}, where WERM, trained with  
the usual gradient descent method (i.e., without clipping or adding noise) \textcolor{black}{ provides 
 relative privacy guarantees---as measured by the success of MIAs---qualitatively aligned with the conclusions of Theorem~\ref{th:privacy-theorem}. } 

\subsection{WERM's Model Utility}\label{sec:utility-analysis}
We provide a bound for 
\textcolor{black}{the expected loss of the model learned through WERM ($f_{\theta_{\scaleto{\text{WERM}}{3.5pt}}}$) with respect to the smallest possible loss $\min_{\theta \in \Theta} L_{\mathbb{D}}(f_{\theta})$.}
\textcolor{black}{
\begin{theorem}[Generalization bound]\label{th:gen-bound}
Under the assumption that the loss function is 
bounded in the range [0, 1], it follows that:
\begin{equation}\label{eq:gen-bound-formal}
    \begin{aligned}
    L_{\mathbb{D}}&(f_{\theta_{\scaleto{\text{WERM}}{3.5pt}}}) \leq \min_{\theta \in \Theta} L_{\mathbb{D}}(f_\theta) \\    
    & + 2 \: \sqrt{\frac{\text{VCdim}(\Theta)}{N_{\text{eff}}}} \: \cdot \: \sqrt{\gamma_{2} + \log\Biggl(\frac{N}{\text{VCdim}(\Theta)}\Biggr)} 
     + \sqrt{\frac{2 \ln 2 / \delta}{N_{\text{eff}}}}
    \end{aligned}
\end{equation}
with probability $\geq 1 - \delta$,  where: 
\begin{align*}
f_{\theta_{\scaleto{\text{WERM}}{3.5pt}}} &= \argmin_{\theta \in \Theta} L^{w}_{D}(f_{\theta}), \\
L^{w}_{D}(f_\theta) &= (1-w) L_{D_{T}}(f_\theta) + w L_{D_{R}}(f_\theta), \\
\gamma_{2} &= \max \Bigl\{\frac{4}{\text{VCdim}(\Theta)}, 1 \Bigr\}, 
N_{\text{eff}} = \left[\frac{(1 - w)^{2}}{N_{T}} + \frac{w^{2}}{N_{R}} \right]^{-1},
\end{align*}
$D_{T}\sim \mathbb D^{N_T}$, $D_{R} \sim \mathbb{D}^{N_R}$, 
$D = D_{T} \cup D_{R}$, $N = |D|$, 
and VCdim($\Theta$) is the VC-dimension of hypothesis class $F_{\Theta} = \{ f_{\theta} : \theta \in \Theta \} $.
\end{theorem}
}

The proof of Theorem~\ref{th:gen-bound} can be found in Appendix~\ref{app:proof-gen-bound}. 
For our purposes, it is important to keep in mind that a smaller bound in~\eqref{eq:gen-bound-formal} implies that the performance of a model learned by WERM will be closer to the performance of the best model in the hypothesis class $F_{\Theta}$, resulting in higher model utility.



Theorem~\ref{th:gen-bound} makes clear how the classifier's utility depends directly on $w$. Given a fixed total dataset size and an already selected model class, $N$ and $\text{VCdim}(\Theta)$ in~\eqref{eq:gen-bound-formal} are held constant. Consequently, the only term that influences the generalization bound is \textcolor{black}{the \emph{effective number of samples}} $N_{\text{eff}}$: the larger $N_{\text{eff}}$, the higher the model's utility. 
It is easy to show that the effective dataset size is always upper-bounded by the total dataset size (i.e., $N_{\text{eff}}\le N$) and is maximized for $w^*= N_R/N$. This choice results in the same weight given to every sample independently of whether it belongs to the training or reference data, i.e., $L^{w}_{D}= \frac{1}{N} \sum_{(x, y) \in D_{T} \cup D_{R}} \ell(f_{\theta}, (x, y))$.
When $w=w^*$, training and reference data have the same privacy guarantees ($\epsilon_T=\epsilon_R$, see Section~\ref{sec:privacy-analysis}). Alternatively, when the privacy considerations are unequal, $N_{\text{eff}}$ degrades quadratically with respect to the difference between $w$ and $w^*$. 
We can thus conclude that heterogeneous privacy requirements for training and reference data (i.e., $\epsilon_{_{T}} \neq \epsilon_{_{R}}$) lead to samples being weighted differently in the two datasets, which causes an increase in the privacy of a selected dataset at the expense of overall model utility.

\subsection{Theoretical Utility-Privacy Tradeoff}\label{sec:theoretical-tradeoff-analysis}
\begin{figure*}
\includegraphics[scale=0.5]{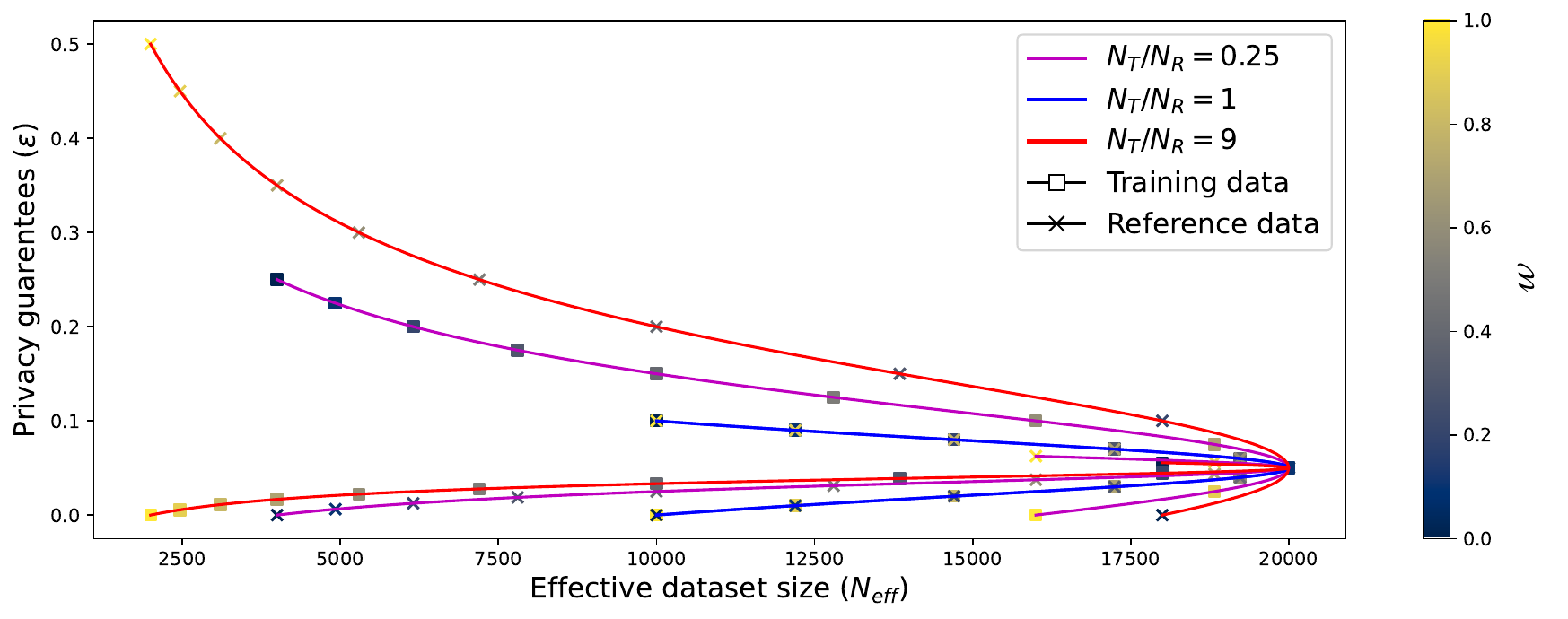}
\caption{\textcolor{black}{Theoretical utility-privacy tradeoff for WERM trained with DP-SGD as derived by the bounds in Theorem~\ref{th:privacy-theorem} and Theorem~\ref{th:gen-bound}, $\mathbf{\epsilon_{_{0}} = 1000}$ and $\mathbf{N=N_T+N_R=20,000}$. 
The curves show the effective dataset size ($\mathbf{N_{\text{eff}}}$, a proxy for model utility), training data privacy guarantees ($\mathbf{\epsilon_{_{T}}}$), and reference data privacy guarantees ($\mathbf{\epsilon_{_{R}}}$) are influenced by reference data weight values ($\mathbf{w}$) and dataset size proportion ($\mathbf{N_T/N_R}$). In Section~\ref{sec:empirical-results}, we show that the theoretical results presented here are aligned with the empirical results presented in Figure~\ref{fig:three-curves}.}}\label{fig:theoretical-bounds}
\Description{The effective dataset size is plotted in comparison to the privacy guarantees for training and reference data to show how the weight term in WERM impacts all of these values.}
\end{figure*}

By combining Theorem~\ref{th:privacy-theorem} and Theorem~\ref{th:gen-bound}, we can study WERM's utility-privacy tradeoff for different dataset sizes and different values of the weight $w$.
\textcolor{black}{In Figure~\ref{fig:theoretical-bounds}, we plot $\epsilon$ privacy values against effective dataset sizes ($N_{\text{eff}}$) for different proportions of training data and reference data ($\frac{N_{T}}{N_{R}}$)
and weight values in the interval $[0.0, 1.0]$ using a fixed total dataset size (N) of 20000. We select values of $\frac{N_{T}}{N_{R}}$ equal to 0.25, 1, and 9 to represent each possible distinct data setting ($N_T > N_R, N_T = N_R, N_T < N_R$) without leading to overlapping curves.}\footnote{\textcolor{black}{We observe that the training (resp.~reference) data curve for a given ratio ${N_{T}}/{N_{R}}=a$ coincides with the reference (resp.~training) data curve for  the reciprocal ${N_{T}}/{N_{R}}={1}/{a}$. Thus, it is also possible to observe the behavior for ${N_{T}}/{N_{R}} = 4$ and ${N_{T}}{N_{R}} = {1}/{9}$ in Figure~\ref{fig:theoretical-bounds}. We show an example for ${N_{T}}/{N_{R}} = 0.25$ in Figure~\ref{fig:inverse-theoretical-curves} (Appendix~\ref{app:additional-figs}).}}

For a given dataset size proportion, by varying the reference data weight, $w$, WERM is capable of achieving a wide spectrum of tradeoffs for model utility, training data privacy, and reference data privacy. When $w=0$, indicated by the darkest colored points in Figure~\ref{fig:theoretical-bounds}, the reference data is fully protected (marked with ``x'), while the training data is most exposed (marked with ``o''). 
As $w$ increases, indicated by the color of the points becoming lighter, sacrificing reference data privacy (i.e., $\epsilon_{_{R}}$ increasing) leads to greater training data privacy protection (i.e., $\epsilon_{_{T}}$ decreasing). 

The interaction between $w$ and model utility is particularly interesting. We observe that increasing the weight causes the model utility to first increase and then decrease. The maximum utility ($N_{\text{eff}}=N$) is obtained for $w^*= N_R/N$, which coincides with the setting of equal privacy guarantees for the two datasets ($\epsilon_{_T}=\epsilon_{_R}=\frac{\epsilon_0}{N}$). This result is independent of the relative size of the two datasets, and indeed we can observe that all curves share the point $(N, \epsilon_0/N)$. \textcolor{black}{When $\epsilon_0$ increases, the $y$-scale in Figure~\ref{fig:theoretical-bounds} increases accordingly, but the shape of the curves does not change.} Overall, Figure~\ref{fig:theoretical-bounds} confirms that WERM's utility-privacy tradeoff is easy to interpret, which is highly desirable for its role as a baseline defense.

\section{Experiments}\label{sec:experiments}
In this section, we outline our training strategy and evaluation setting, describe in detail our process for training each empirical privacy defense, and conduct a systematic evaluation of our WERM baseline in a variety of utility-privacy settings. \textcolor{black}{Ultimately, we will demonstrate that WERM's empirical utility-privacy tradeoff (Figure~\ref{fig:three-curves}) is qualitatively similar to what is predicted by the theoretical analysis (Figure~\ref{fig:theoretical-bounds}),
which confirms our intuition that WERM is an interpretable baseline in both theory and practice.}

\subsection{Datasets}
We chose to conduct our experiments on the Purchase100, Texas100, and CIFAR100 datasets because they have been widely used for assessing empirical privacy defenses and MIAs~\cite{nasr2018machine,nasr2019comprehensive,yeom2018privacy,song2021systematic,li2021membership,shejwalkar2021membership,yang2020defending,jia2019memguard,shokri2017membership}. A detailed description of each dataset is provided in Appendix~\ref{dataset-desc}.

\subsection{Methodology}

\subsubsection{Training}\label{sec:training-strategy}
Conducting a fair comparison of empirical privacy defenses requires using a standardized approach for dataset pre-processing (e.g., equivalent training/reference/test data size proportions) and model architecture choices for all methods. As AdvReg~\cite{nasr2018machine} is the first proposed empirical privacy defense and most well-studied method, its experimental setting has consequently become the de-facto standard for comparing defenses on the Purchase100 and Texas100 dataset~\cite{song2021systematic,li2021membership,choquette2021label}. In this setting, one applies the defense mechanism to a 4-layer fully connected neural network classifier with layer sizes [1024, 512, 256, 100], and uses 10\% of Purchase100 ($\approx$ 20,000 samples) and 15\% of Texas100 ($\approx$ 10,000 samples) as training data. For the CIFAR100 dataset, we use 20,000 samples for training data and align our study with more recent evaluations that consider a ResNet-18~\cite{he2016deep} as the classification model~\cite{li2021membership,wang2020against}. We assume that each defense has access to reference data that is the same size as the training data. Following the strategy of the original AdvReg experiments, all classification models are trained using an Adam optimizer~\cite{kingma2014adam} with a learning rate equal to 0.001. For reasons that are described in Appendix~\ref{app:mmd}, MMD requires using a batch size equal to 512 or greater. This contrasts with the original AdvReg experiments that use a batch size equal to 128. To ensure a fair comparison, we train all evaluated defenses using both batch sizes and select the best version for each method. For a given defense, the reported results are mean values over 10 training runs for different seeds of a random number generator. \textcolor{black}{Following the same training strategy as previous works~\cite{nasr2018machine,song2021systematic,choquette2021label}, we train each defense for a specific number of epochs that ensures the model converges without severely overfitting.\footnote{\textcolor{black}{It is also possible to use validation data to find an opportune epoch to end training.
However, using a validation dataset introduces questions regarding the validation data’s degree of privacy leakage~\cite{papernot2021hyperparameter}. As we are already evaluating the relative training and reference data privacy leakage, introducing another dataset will add further complexity to our analysis, which will make it more difficult to interpret the results. 
In Figure~\ref{fig:three-curves-valid} (Appendix~\ref{app:additional-figs}), we present the utility-privacy curves using validation data to determine the number of training epochs. The difference is negligible compared to our results using a predetermined number of epochs.}}
Additionally, the regularization values we select for training each defense are explicitly chosen to demonstrate all possible relative privacy levels that a given method can achieve.\footnote{
\textcolor{black}{As discussed in Section~\ref{sec:empirical_privacy_defenses}, higher values of the regularization value should lead to higher privacy protection for training data, potentially leaking more information about reference data.}
}}

\subsubsection{Evaluation}\label{sec:eval-method}
\textcolor{black}{We use the same methodology and released code\footnote{https://github.com/inspire-group/membership-inference-evaluation} as Song and Mittal~\cite{song2021systematic}, where an empirical privacy defense is evaluated against three threshold-based MIAs and the gap attack~\cite{yeom2018privacy}. Additionally, we evaluate against a neural network-based attack~\cite{shokri2017membership} that could be executed by a stronger adversary with access to training/non-training samples, and the results, shown in Figure~\ref{fig:three-curves-nn} (Appendix~\ref{app:additional-figs}), are qualitatively similar to those in Figure~\ref{fig:three-curves}.}

\textcolor{black}{Following the standard evaluation methodology~\cite{nasr2018machine,li2021membership,song2021systematic,jia2019memguard,choquette2021label}, a distinct test dataset from the same underlying distribution as the training data is used to evaluate the final accuracy of the trained model and as the ``non-training'' data to evaluate (together with part of the training data) the accuracy of the MIA according to~\eqref{mi-attack-acc}.} Across all datasets, the two most effective attacks were threshold-based and used either the confidence value or modified-entropy. \textcolor{black}{To quantify the privacy leakage} in our experimental results, we report the MIA accuracy of the attack using confidence values because it requires less assumptions and performs equivalently well.

In our evaluation, we explicitly measure the capabilities of empirical privacy defenses in a variety of model utility, training data privacy, and reference data privacy settings to determine the most effective methods in each case. We define the notion of a \emph{model instance} as a defended classifier obtained by training with a certain regularization or weight value. This terminology will be used as we select model instances that most closely adhere to a specific privacy setting (e.g., WERM trained with a weight value $w=0.5$ is a model instance that is coherent with equal training and reference data privacy requirements, as the two datasets have the same size).

\subsection{Evaluated Defenses}
As WERM is designed to be a baseline for empirical privacy defenses using reference data, we only compare against methods in this category, which excludes the recently proposed Self-Distillation~\cite{tang2021mitigating}. Specifically, we evaluate AdvReg~\cite{nasr2018machine}, which is the most well-studied, and MMD~\cite{li2021membership}, which is the current state-of-the-art. We do not consider confidence-vector masking defenses~\cite{jia2019memguard,yang2020defending} because they have been shown to be ineffective against label-based attacks such as the simple gap attack~\eqref{gap-attack-single}~\cite{song2021systematic,choquette2021label}.

\subsubsection{WERM}
\textcolor{black}{Using the classification models described in Section~\ref{sec:training-strategy} for each dataset, we train WERM using weight values equal to 0.0, 0.03, 0.1, 0.3, and 0.5, as well as values of  0.98 and 0.02 (Purchase100), 0.999 and 0.001 (Texas100), and 0.9975 and 0.005 (CIFAR100) that are chosen specifically to achieve the constraints for ``public reference data'' and ``high reference data privacy'' as outlined in Section~\ref{sec:empirical-results}}.\footnote{\textcolor{black}{Due to the symmetric role of training and reference data in WERM, privacy evaluation for training data for a given value $w$ corresponds to privacy evaluation for reference data for $1-w$. In practice, the reported results therefore allow for evaluating a larger range of values including 0.7, 0.9, 0.97, and 1.0 for all datasets and 0.98, 0.999, and 0.995 for Purchase100, Texas100, and CIFAR100, respectively.}} These weights were chosen to reflect the full range of utility-privacy tradeoffs that the method can achieve. To train WERM, for all reference data weight configurations, we fix the number of training epochs at 20, 4, and 25 for the Purchase100, Texas100, and CIFAR100 datasets, respectively. \textcolor{black}{The number of training epochs for WERM, as well as for the other empirical privacy defenses we evaluate, are selected based on the standard methodology discussed in Section~\ref{sec:training-strategy}. Additionally, in all our experiments, we use the standard version of gradient descent, and the resulting models, therefore, have no formal DP guarantees. Nevertheless, we show that WERM's relative privacy guarantees---as measured by  MIA accuracy---qualitatively align with the conclusions of Theorem~\ref{th:privacy-theorem}, a result that is justified by our discussion at the end of Section~\ref{sec:privacy-analysis}. }

While we analyze the generalization bound of WERM in Section~\ref{sec:utility-analysis} for the setting where the empirical loss is minimized, in practice it is possible to end training before convergence. This simple technique, known as early stopping~\cite{caruana2000overfitting}, has been observed to protect privacy~\cite{song2021systematic}. 
As WERM can potentially benefit from early stopping without incurring a loss of interpretability, we evaluate a version of our baseline, henceforth referred to as WERM-ES, that uses this approach. To train WERM-ES, for all reference data weight configurations, we fix the number of training epochs at 7, 1, and 6 for the Purchase100, Texas100, and CIFAR100 datasets, respectively. 

\subsubsection{Adversarial Regularization}\label{sec:adv-reg-exp}
Our AdvReg implementation relies on the officially released code\footnote{https://github.com/SPIN-UMass/ML-Privacy-Regulization} with a few changes to solve several problems we discuss in Appendix~\ref{app:adv-reg}. 



We also evaluate a variant of AdvReg that can be obtained by modifying the gradient update in~\cite{nasr2018machine}. 
Although the declared objective is to solve problem~\eqref{adv-reg-opt}, 
when taking the gradient of~\eqref{adv-reg-opt} with respect to $\theta$, Nasr et al.~\cite{nasr2018machine} only consider the terms evaluated on training data:
\begin{equation}
    \nabla_{\theta} \frac{1}{m} \sum_{i=1}^{m} \ell(f_{\theta}, (x_{i}, y_{i})) + \lambda \: \log[h_{\phi}(x_{i}, y_{i}, f_{\theta}(x_{i}))]
\end{equation}
However, the gradient of~\eqref{adv-reg-opt} with respect to $\theta$ contains an additional term that is evaluated on the reference data: 
\begin{equation}\label{non-member-term}
\frac{\lambda}{m^{\prime}}\sum_{i=1}^{m^{\prime}} \log[1 - h_{\phi}(x_{i}^{\prime}, y_{i}^{\prime}, f_{\theta}(x_{i}^{\prime}))]
\end{equation}
We refer to this variant using the reference data term as AdvReg-RT.
As was observed in Figure~\ref{fig:adv_reg-tradeoff}, AdvReg-RT achieves a distinct set of model utility, training data privacy, and reference data privacy tradeoffs compared to AdvReg. We therefore choose to compare both formulations with our WERM baseline.

Using the classification models described in Section \ref{sec:training-strategy} for each dataset, we train both versions of AdvReg using regularization values equal to 1, 2, 3, 6, 10, and 20 for Purchase100 and Texas100 and 1e-6, 1e-3, 1e-1, and 1 for CIFAR100. These values were selected on a per dataset basis to best represent the utility-privacy tradeoff that each formulation is capable of achieving.  The number of training epochs is fixed at 10, 10, and 25 when training AdvReg and 35, 20, and 25 when training AdvReg-RT, for the Purchase100, Texas100, and CIFAR100 datasets, respectively. 

\subsubsection{MMD-based Regularization}\label{sec:mmd-explanation}
Using the classification models described in Section \ref{sec:training-strategy} for each dataset, 
we train MMD using regularization values equal to 
0.1, 0.2, 0.35, 0.7, and 1.5 that demonstrate the total achievable utility-privacy curve. As the released code implementation of AdvReg~\cite{nasr2018machine} benefits from training the classifier for a few warm-up steps without regularization, we also train MMD with and without a warm-up, reporting only the best results for each dataset. The number of training epochs is fixed at 25, 8, and 15 when training without warm-up steps and 20, 8, and 8 when training with warm-up steps, for the Purchase100, Texas100, and CIFAR100 datasets, respectively. Details about the implementation can be found in Appendix~\ref{app:mmd}.

\subsection{Empirical Results}\label{sec:empirical-results}
\begin{figure*}
     \centering
     \includegraphics[width=1\linewidth]{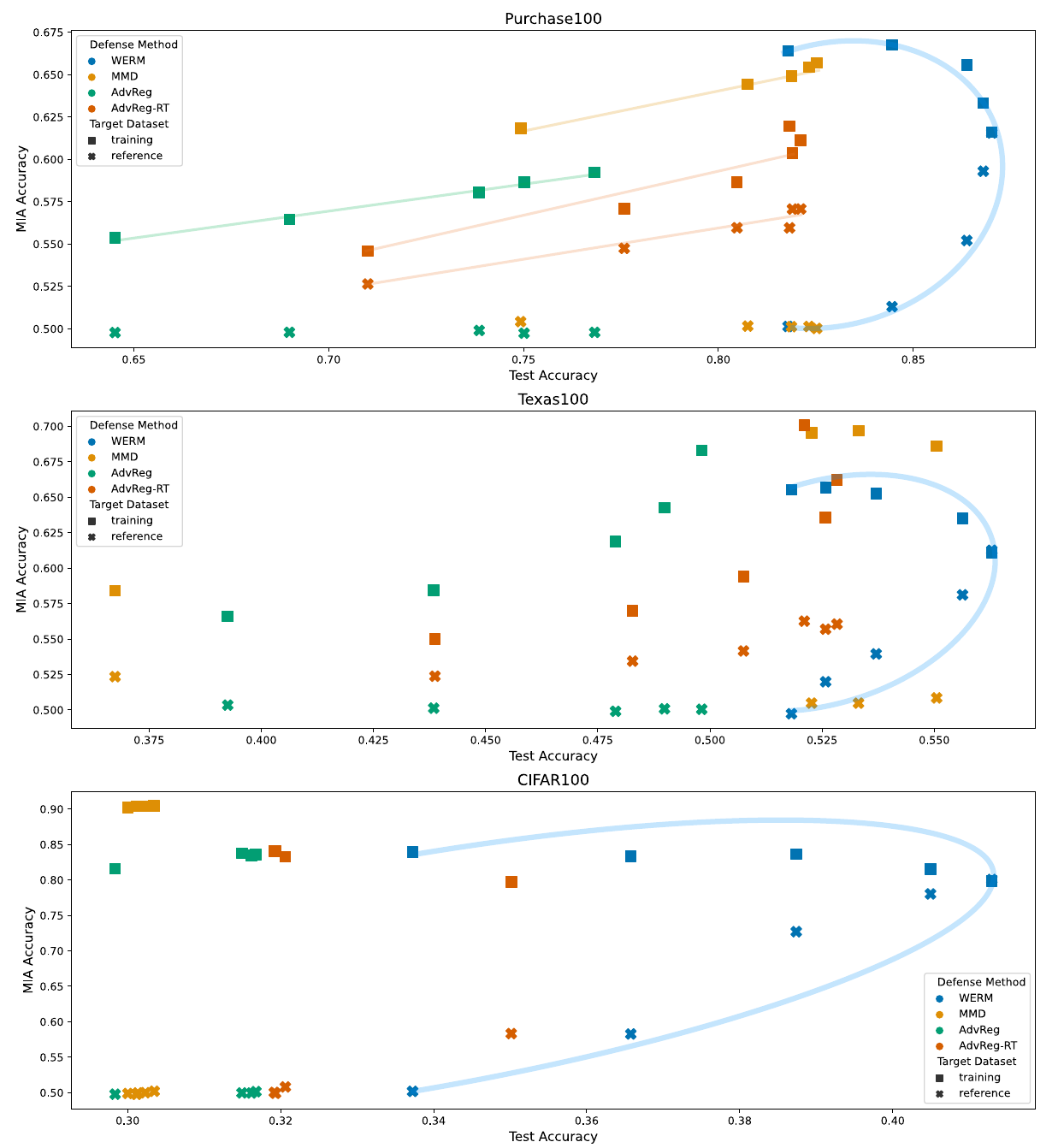}
    \caption{Utility-privacy tradeoffs obtained by various empirical privacy defenses for the Purchase100, Texas100, and CIFAR100 datasets. The test accuracy of a defended classifier is measured using unseen test data and the MIA accuracy on training and reference data is evaluated with a threshold-based using confidence values (Eq.~\ref{thresh-attack}). Each point on the curve represents the evaluation of a model instance using a distinct regularization value (for AdvReg, AdvReg-RT, and MMD) and reference data weight value (for WERM). \textcolor{black}{We highlight some qualitative trends to help demonstrate that the empirical curves coincide with the theoretical curves in Figure~\ref{fig:theoretical-bounds}.}}
    \label{fig:three-curves}
    \Description{The utility-privacy curves are plotted for all of the empirical privacy defenses that we evaluate and compared against our WERM baseline. In the first plot of the figure associated to the Purchase100 dataset, we highlight the curve that our method generates with a half-moon shape, whereas the other methods have linearly changing privacy leakages.}
\end{figure*}

In Figure~\ref{fig:three-curves}, we show the empirical utility-privacy tradeoffs obtained by AdvReg~\cite{nasr2018machine}, MMD~\cite{li2021membership}, AdvReg-RT, and WERM for the Purchase100, Texas100, and CIFAR100 datasets. \textcolor{black}{In these plots, we show the exact points that make up the curve, as well as some qualitative lines to highlight the trends and improve readability. The curves derived from theoretical bounds in Figure~\ref{fig:theoretical-bounds} and from experimental results in Figure~\ref{fig:three-curves} both show utility vs. privacy. However, Figure~\ref{fig:theoretical-bounds} evaluates the utility through the effective number of samples, $N_{\text{eff}}$, and \textcolor{black}{privacy leakage through the DP parameters, $\epsilon_{_{T}}$ and $\epsilon_{_{R}}$},
whereas Figure~\ref{fig:three-curves} uses the test accuracy and MIA accuracy.}
Table~\ref{tab:results-table} focuses specifically on our three key privacy settings: \textcolor{black}{public reference data,} equal training-reference data privacy and high reference data privacy.

\begin{table*}
\caption{Comparison of test accuracy and MIA accuracy for AdvReg, AdvReg-RT, MMD, WERM, and WERM-ES under the settings of \textcolor{black}{public reference data,} equal privacy considerations, and high reference data privacy. A dashed line (---) means that the defense produced no model instances that met the criteria. The values under ``MIA Train'' and ``MIA Ref'' represent the membership inference attack accuracy on training and reference data, respectively.}
\centering
\small
\begin{tabular}{c c ccc ccc ccc}
    \toprule
\multirow{2}{*}{Dataset} & \multirow{2}{*}{Defense}
        & \multicolumn{3}{c}{Public Reference Data} & \multicolumn{3}{c}{Equal Privacy Considerations} & \multicolumn{3}{c}{High Reference Data Privacy}   \\
    \cmidrule(lr){3-5} \cmidrule(lr){6-8} \cmidrule(lr){9-11}
         & & Test Acc. & MIA Train & MIA Ref & Test Acc. & MIA Train & MIA Ref & Test Acc. & MIA Train & MIA Ref  \\
    \midrule
\multirow{4}{*}{Purchase100}  
        & AdvReg        & ---       & ---       & --- & ---       & ---       & ---       & 76.8\%    & 59.2\%    & 50.0\%                \\
        & AdvReg-RT     & ---       & ---       & --- & 82.1\%    & 61.1\%    & 57.1\%    & ---       & ---       & ---           \\
        & MMD           & ---       & ---       & --- & ---       & ---       & ---       & 82.5\%    & 65.7\%    & 50.0\%                \\
        & WERM  & 83.8\%       & 51.0\%      & 66.7\%  & 87.0\%    & 61.5\%    & 61.5\%    & 84.2\%    & 68.0\%    & 50.9\%        \\
        & WERM-ES  & 77.9\%       & 50.1\%     & 57.4\% & 83.6\%    & 54.7\%    & 54.9\% & 78.4\%       & 57.8\%       & 50.0\%         \\
    \addlinespace
    \hline
    \addlinespace
\multirow{4}{*}{Texas100}
        & AdvReg        & ---       & ---       & --- & ---       & ---       & ---       & 49.8\%    & 68.3\%    & 50.0\%                \\
        & AdvReg-RT     & ---       & ---       & --- & 48.3\%    & 57.0\%    & 53.4\%    & ---       & ---       & ---           \\
        & MMD           & ---       & ---       & --- & ---       & ---       & ---       & 55.1\%    & 68.6\%    & 50.8\%              \\
        & WERM  & 54.2\%      & 50.9\%       & 65.6\% & 56.3\%    & 61.1\%    & 61.3\%    & 52.0\%    & 65.7\%    & 50.9\%       \\
        & WERM-ES  & 43.7\%      & 50.3\%       & 55.9\% & 49.8\%    & 54.3\%    & 54.0\% & 44.4\% & 56.3\% & 50.5\%         \\
    \bottomrule
    \addlinespace
\multirow{4}{*}{CIFAR100}
        & AdvReg         & ---       & ---       & --- &---       & ---       & ---        & 31.7\%    & 83.6\%    & 50.0\%              \\
        & AdvReg-RT     & ---       & ---       & --- & ---       & ---       & ---       & 31.1\%    & 83.3\%    & 50.8\%               \\
        & MMD           & ---       & ---       & --- & ---       & ---       & ---       & 30.2\%    & 90.4\%    & 50.0\%               \\
        & WERM  & 34.2\%       & 50.5\%       & 84.0\%  & 41.3\%    & 79.8\%    & 80.1\%    & 33.8\%    & 83.5\%    & 50.9\%        \\
        & WERM-ES  & 32.9\%       & 50.1\%       & 63.3\%  & 40.1\%    & 60.2\%    & 60.2\% & 33.0\% & 63.6\%    & 50.0\%       \\
    \bottomrule
    \addlinespace
\end{tabular}
\label{tab:results-table}
\end{table*}

\subsubsection{Utility-Privacy Curve Analysis}\label{sec:utility-privacy-curve-analysis}
The three objectives of model utility, training data privacy leakage, and reference data privacy leakage are inherently in conflict with one another. Ideally, one would like to see that an empirical privacy defense can produce a landscape of model instances that spans a vast range of utility-privacy regimes. In theory, each of the methods we evaluate should have this capability, as they all have a mechanism for controlling the amount of regularization that is applied during training. Examining the utility-privacy curves in Figure~\ref{fig:three-curves} allows us to understand the tradeoffs that the various defenses can achieve in practice. As noted in Section~\ref{sec:eval-method}, we quantify privacy leakage using a threshold-based MIA on a classifier's confidence values.\footnote{\textcolor{black}{A random guesser would get an average expected accuracy of 0.5 but its average accuracy on a finite dataset can either exceed or fall short of 0.5. It should then not be surprising that some attacks have an accuracy marginally below 0.5. (e.g., 0.498), as can be seen in Figure~\ref{fig:three-curves}.}}

WERM is the only defense that can clearly tradeoff between the three objectives. For Purchase100, over the range of privacy settings from equal privacy ($w=0.5$) to high reference data privacy ($w=0$), we see that WERM achieves values of 87\% / 54.7\% / 54.9\% and 81.8\% / 57.6\% / 50\%, for test accuracy (model utility), MIA accuracy on training data, and MIA accuracy on reference data, respectively. Between these edge cases, we see that WERM can produce model instances capable of trading off reference data privacy for both model utility and training data privacy. The same trend can be observed for WERM on the Texas100 and CIFAR100 datasets. 
Regarding WERM-ES, as shown in Figure~\ref{fig:three-curves-full} (Appendix~\ref{app:additional-figs}), the defense exhibits equivalent behavior to WERM, but, as expected, it achieves higher overall privacy protection at the cost of lower 
model utility. Figure~\ref{fig:three-curves-full} also contains the utility-privacy curves for early stopping (EarlyStop) using only the training data. 

Looking at the state-of-the-art defenses we evaluate (AdvReg-RT, AdvReg, and MMD) reveals two situations. First, we can see that AdvReg-RT is able to sacrifice model utility for overall better privacy protection. However, it does not have the ability to trade off between training data privacy and reference data privacy, as changing the regularization value $\lambda$ results in the
training and reference data privacy leakage increasing/decreasing together. Due to this limited functionality, AdvReg-RT is never able to reach the setting where training data privacy protection is equal to reference data privacy protection. Second, the utility-privacy curves for AdvReg and MMD demonstrate that these defenses are completely unable to trade reference data privacy for either model utility or training data privacy. While training data privacy can be sacrificed for better model utility, attack accuracy on reference data never materially changes, remaining below 51\% for both defenses at all meaningful test accuracy values. 
Overall, we observe that for the entire curve of possible utility-privacy tradeoffs, excluding the high reference data privacy setting, WERM/WERM-ES is unequivocally the best-performing method, and in any regime where training data privacy is valued absolutely equal to or greater than reference data privacy (including the public reference data case), WERM/WERM-ES is, in fact, the only viable defense. 

\textcolor{black}{In addition to WERM being a baseline defense with good utility, we also want its output to align with the desired relative privacy level that is encoded in a given choice of $w$.}
Comparing WERM's empirical utility-privacy tradeoffs in Figure~\ref{fig:three-curves} with the theoretical tradeoffs in Figure~\ref{fig:theoretical-bounds}, we see that they exhibit the same trend where a gradual transition occurs from the setting of high reference data privacy to that of equal privacy  over the weight value interval of [0.0, 0.5]. \textcolor{black}{The fact that our theoretical bounds are qualitatively aligned with our experimental results helps to demonstrate that WERM is indeed an interpretable baseline. We conduct a quantitative comparison in Section~\ref{sec:weight-term-selection}.}

\subsubsection{Public Reference Data}
\textcolor{black}{First, we examine the case where reference data is public. In this setting, the privacy of reference data is of no concern. Therefore, an optimal defense should utilize the reference data to the furthest extent possible to decrease training data privacy leakage and increase test accuracy. For a given empirical privacy defense, we select model instances using the following procedure: 1) Identify all model instances with MIA accuracy on training data less than or equal to 51\%, 2) Among the model instances meeting this criterion, select the one with best test accuracy. As can be observed in Table~\ref{tab:results-table}, on Purchase100, Texas100, and CIFAR100, only WERM or WERM-ES are able to produce suitable model instances in this setting. Although MMD and AdvReg include a regularization term that is intended to tradeoff  privacy against test accuracy, the methods are simply not capable of maximally exploiting reference data. Alternatively, WERM can sacrifice reference data privacy to achieve a high test accuracy and strict training data privacy protection.}

\subsubsection{Equal Privacy Requirements}\label{sec:equal-privacy}
\textcolor{black}{Second, we examine the setting where training and reference data have equal privacy requirements.} For a given empirical privacy defense, we select the model instance using the following procedure: 1) Identify all model instances where the difference between the attack accuracy on training data and reference data is less than or equal to 4\%, 2) Among the model instances meeting this condition, select the one with best test accuracy. We use 4\% as the threshold to define ``equal'' privacy considerations because at lower thresholds AdvReg-RT is not able to achieve a model instance with satisfactory utility to be relevant for comparison. Table~\ref{tab:results-table} shows that only WERM/WERM-ES and AdvReg-RT are able to operate in this privacy regime; MMD and AdvReg fail to produce any viable model instances.

On Purchase100 and Texas100, for the selected model instances, WERM-ES outperforms AdvReg-RT on all three objectives. Compared to AdvReg-RT and WERM-ES, WERM achieves significantly higher model utility, at the expense of worse training and reference data privacy. On CIFAR100, AdvReg-RT is unable to yield a model instance that meets the conditions, making WERM/WERM-ES the only working defense. 

\subsubsection{High Reference Data Privacy}\label{sec:extreme-ref-privacy}
Lastly, we examine the case where reference data is considered highly private and its privacy can therefore only be minimally sacrificed. For a given empirical privacy defense, we select model instances using the following procedure: 1) Identify all model instances with MIA accuracy on reference data less than or equal to 51\%, 2) Among the model instances meeting this criterion, select the one with best test accuracy. 
In Table~\ref{tab:results-table}, it can be seen that on Purchase100 and Texas100, AdvReg, MMD, WERM, and WERM-ES are all able to produce a model instance that leaks only minimal reference data privacy according to our selection method,
whereas AdvReg-RT is unable to yield a suitable model instance. However, on CIFAR100, all five defenses achieve a valid result.

\textcolor{black}{By setting very strict privacy requirements for reference data, we aim to remove one objective from the evaluation such that we can make a comparison based solely on the model utility and training data privacy leakage. Nonetheless, it is still possible that two model instances satisfy the reference data privacy requirement but cannot be definitively compared because one can achieve better utility and the other better training data privacy protection. On Purchase100, for example, WERM-ES outperforms AdvReg, but it is not clear which method is preferable between WERM, WERM-ES, and MMD without deciding the relative importance assigned to model utility and training data privacy. WERM can achieve higher test accuracy, WERM-ES can achieve higher privacy protection on training data, and MMD can achieve a utility-privacy tradeoff that is a middle ground between these two methods. On Texas100, WERM and MMD both outperform AdvReg, but a comparison between the two also requires making explicit the relative importance of model utility and training data privacy. On CIFAR100, however, WERM/WERM-ES is clearly superior to the other three defenses.}


\section{Discussion}\label{sec:discussion}
\subsection{\textcolor{black}{Selection of Defense Parameters
}}\label{sec:weight-term-selection}
\begin{table}
\normalsize
\caption{\textcolor{black}{Comparison of the Pearson Correlation Coefficient between the training-reference data desired privacy ratio (as determined by the choice of $w$ or $\lambda$) and the empirical privacy ratio (as measured by a MIA) for WERM, AdvReg, and MMD. The coefficient is computed across the Purchase100, Texas100, and CIFAR100 datasets.}}\label{table:alignment-analysis}
\centering
\begin{tabular}{c c c c }
    \toprule
Defense &  WERM & AdvReg & MMD \\
    \midrule
    Pearson Correlation Coefficient        & 0.84 & 0.07 & 0.48 \\
    \addlinespace
    \hline
\end{tabular}
\end{table}
\textcolor{black}{Even when a model training entity has clearly defined its desired relative privacy level between training and reference data, realizing a classifier with this exact degree of relative privacy protection still requires selecting the corresponding parameters of the empirical privacy defense: the reference data weight term ($w$) for WERM and regularization weight term ($\lambda$) for AdvReg or MMD.
As an illustration, if the two datasets are of equal size, and the reference data needs to be twice as private, what should be the chosen values of $w$ for WERM or $\lambda$ for AdvReg or MMD?
We argue that an empirical privacy defense becomes more practical if there exists an intelligible (e.g., linear) mapping to guide a machine learning practitioner in the selection of a defense parameter.
AdvReg and MMD do not provide a practical guideline except for the general intuition that a larger value of $\lambda$ should provide higher training data privacy and lower reference data privacy.
For WERM, the parameter $w$ can be adjusted to ensure a specified theoretical level of relative privacy, as dictated by the equations in Theorem~\ref{th:gen-bound} (${\epsilon_T}/{\epsilon_R} = {(1-w)}/{w} \times {N_R}/{N_T}$). 
However, translating DP-like theoretical privacy guarantees into practical privacy guarantees e.g., in terms of MIA accuracy, is a highly complex and still unresolved issue in the field of privacy-preserving machine learning~\cite{carlini2022membership,bernau2019assessing,nasr2021adversary,ye2021enhanced}.
The effectiveness of such a configuration rule therefore needs to be evaluated in terms of the empirical privacy leakage.
}

\textcolor{black}{
In Table~\ref{table:alignment-analysis} we report the Pearson correlation coefficient (PCC) between WERM's theoretical relative privacy, as defined by the value $\epsilon_T/\epsilon_R$, and its empirical relative privacy, as measured by the ratio between the MIA accuracy on training data and on reference data. The coefficient gauges the linear correlation between the two quantities.
For AdvReg and MMD, 
without clear configuration guidelines, we report the PCC between the reciprocal of the regularization parameter $1/\lambda$ and the empirical relative privacy. WERM displays the largest PCC at 0.84, which stands in stark contrast to the 0.07 for AdvReg and 0.48 for MMD. These results underscore  WERM as the sole method offering a practical configuration rule to achieve a target relative privacy. 
}

\subsection{Computational Cost Comparison}\label{sec:training-time}
\textcolor{black}{In addition to comparing the utility-privacy tradeoff \textcolor{black}{and practical usability} of empirical
privacy defenses, it is also important to consider their computational
cost. Table~\ref{table:comp-cost} shows, for a fixed batch size, the number of seconds it takes to train each defense for a single epoch and the overall training time considering the total number of epochs used in our experiments. 
We calculate the overall training time as the per epoch training time multiplied by the total number of epochs. Each of these experiments are run using a single GPU on an NVIDIA DGX system. 
Analyzing Table~\ref{table:comp-cost} confirms 
that WERM is indeed significantly less computationally expensive than AdvReg and MMD, as discussed in Section~\ref{sec:method}. On Purchase100, Texas100, and CIFAR100, WERM is 19x, 7x, and 19x faster to train on a per epoch basis, compared to the second fastest method.}

\begin{table}
\caption{\textcolor{black}{Comparison of per epoch and overall training time, in seconds (s), for each empirical privacy defense on Purchase100, Texas100, and CIFAR100.}}\label{table:comp-cost}
\centering
\begin{tabular}{c c c c}
    \toprule
Dataset & Defense & Per Epoch (s) & Overall (s) \\
    \midrule
\multirow{3}{*}{Purchase100}  
        & WERM      & 0.5       & 10    \\
        & AdvReg    & 9.5       & 95    \\
        & MMD       & 16.4      & 328   \\
    \addlinespace
    \hline
    \addlinespace
\multirow{3}{*}{Texas100}
        & WERM      & 0.9       & 3.6   \\
        & AdvReg    & 6.4       & 64    \\
        & MMD       & 6.8       & 54.4  \\
    \bottomrule
    \addlinespace
\multirow{3}{*}{CIFAR100}
        & WERM      & 4.1       & 102.5 \\
        & AdvReg    & 78.8      & 1970  \\
        & MMD       & 94.7      & 757.6   \\
    \bottomrule
    \addlinespace
\end{tabular}
\end{table}

\section{Conclusion and Future Work}\label{sec:conclusion}
In this work, we have analyzed the role of reference data in empirical privacy defenses and identified the issue that reference data privacy leakage must be explicitly considered to conduct a meaningful evaluation. We advanced the current state-of-the-art by proposing a generalization error constrained ERM, which can in practice be evaluated as a weighted ERM over the training and reference datasets. As WERM is intended to function as a baseline, we derive theoretical guarantees about its utility and privacy to ensure that its results will be well-understood 
in all utility-privacy settings. We present experimental results showing that our principled baseline outperforms the most well-studied and current state-of-the-art empirical privacy defenses in nearly all privacy regimes \textcolor{black}{(i.e., independent of the nature of reference data and its level of privacy).} Our experiments also reveal that existing methods are unable to trade off reference data privacy for model utility and/or training data privacy, and thus cannot operate outside of the highly private reference data case. 

\textcolor{black}{Regarding ethical concerns, our proposed baseline operates on the defense side of machine learning privacy; no novel attack has been proposed. 
Nevertheless, our experiments have analyzed the average privacy leakage over the whole dataset, but privacy protection is not always fair across groups in a
dataset~\cite{kulynych2019disparate,de2023empirical}. Future work can evaluate then the fairness of various defense mechanisms using reference data or propose the creation of privacy defenses intended to operate in use-case dependent settings. We hope that our work will continue to motivate the development of a robust evaluation framework for privacy defenses.
}

\begin{acks}
This research was supported in part by ANRT in the framework of a CIFRE PhD (2021/0073) and by the Horizon Europe project dAIEDGE.
\end{acks}


\bibliographystyle{ACM-Reference-Format}
\bibliography{references}

\appendix

\section{WERM Theoretical Analysis}
\subsection{Details on Lagrangian}\label{app:lagrangrian}
The Lagrangian of our constrained minimization problem has the following form:
\begin{equation}\label{w-erm-lagrangian}
    \min_{\theta, \lambda, \mu} \: L_{D} + \lambda \Bigl[ L_{\mathbb{D}} - L_{D_{T}} - c_{T} \Bigr] + \: \mu \Bigl[ L_{\mathbb{D}} - L_{D_{R}} - c_{R} \Bigr].
\end{equation}
Assuming we know the optimal multipliers, $\lambda^{\ast}$ and $\mu^{\ast}$, we can find $\theta$ by minimizing the following equation:
\begin{equation}\label{w-erm-optimal-mults}
    \min_{\theta} \: L_{D} + \lambda^{\ast} \Bigl[ L_{\mathbb{D}} - L_{D_{T}} - c_{T} \Bigr] + \: \mu^{\ast} \Bigl[ L_{\mathbb{D}} - L_{D_{R}} - c_{R} \Bigr]
\end{equation}
or equivalently, as $c_{T}$ and $c_{R}$ are constants:
\begin{equation}\label{w-erm-optimal-mults-no-c}
    \min_{\theta} \: L_{D} + \lambda^{\ast} \Bigl[ L_{\mathbb{D}} - L_{D_{T}} \Bigr] + \: \mu^{\ast} \Bigl[ L_{\mathbb{D}} - L_{D_{R}} \Bigr].
\end{equation}
In the ideal case, one would have access to the underlying distribution, $\mathbb{D}$, and compute all terms in~\eqref{w-erm-optimal-mults-no-c}. Since $\mathbb{D}$ is unknown, we must instead use an approximation. Given that training and reference data are both drawn from $\mathbb{D}$, we estimate $L_{\mathbb{D}}$ as $L_{D_{R}}$ in the second term of~\eqref{w-erm-optimal-mults-no-c} and $L_{\mathbb{D}}$ as $L_{D_{T}}$ in the third term of~\eqref{w-erm-optimal-mults-no-c}. Thus, our formulation becomes:
\begin{align}
    \min_{\theta} \: & L_{D} + \lambda^{\ast} \Bigl[ L_{D_{R}} - L_{D_{T}} \Bigr] + \: \mu^{\ast} \Bigl[ L_{D_{T}} - L_{D_{R}} \Bigr] \label{w-erm-final-form}\\
    & = \: \Bigl[\frac{1}{2} - \lambda^{\ast} + \mu^{\ast} \Bigr] L_{D_{T}} + \: \Bigl[\frac{1}{2} + \lambda^{\ast} - \mu^{\ast} \Bigr] L_{D_{R}} \nonumber.
\end{align}

Although both $\lambda^{\ast}$ and $\mu^{\ast}$ appear in~\eqref{w-erm-final-form}, at the optimum $(\theta^{\ast}, \lambda^{\ast}, \mu^{\ast})$ we expect that only the stricter of the two privacy constraints in~\eqref{w-erm-formulation} will be active (i.e., satisfied with equality) and then with a positive Lagrange multiplier, while the other will be inactive (i.e., it will be a strict inequality) with a null Lagrange multiplier.
Since $D_{T}$ and $D_{R}$ are drawn from $\mathbb{D}$ and have the same size, $L_{D_{T}}$ and $L_{D_{R}}$ are two random variables with the same distribution. As a result, the stricter constraint is determined by the smaller of the two constants ($c_{T}$ and $c_{R}$),
and we will have either $\lambda^{\ast} = 0$ and $\mu^{\ast} > 0$ (if $c_{R} < c_{T}$) or $\lambda^{\ast} > 0$ and $\mu^{\ast} = 0$ (if $c_{T} < c_{R}$). 

{\color{black}
\subsection{Extension to Multiple Datasets}
\label{app:multiple_datasets}
In this paper, we have considered the case when there are two datasets (training data and reference data) with different levels of privacy. We will prove Theorem~\ref{th:privacy-theorem} and Theorem~\ref{th:gen-bound} under the more general case, when there are multiple datasets coming from the same distribution, and each individual dataset can have its own distinct weighting according to its privacy level  (up to the limit case where every point is a separate dataset with distinct privacy).



In this case $L^{w}_{D}$ consists of any possible weighted combination of losses evaluated over multiple datasets, i.e., $L^{w}_{D} = \sum_{m=1}^{M} w_{m} L_{D_{m}}$, and $\sum_{m=1}^{M} w_{m} = 1$. A given dataset, $D_{m}$, is comprised of $|D_{m}|=N_{m}$ data points that are drawn i.i.d. from $\mathbb{D}$, such that $D_{m} = \{z_{m}^{(i)}, i \in [N_{m}]\}$. Accordingly, a given aggregated dataset, $D$, is comprised of $N = |D|$ data points where $N = \sum_{i=1}^{M} N_{m}$.
}

\subsection{Proof of Theorem~\ref{th:privacy-theorem}}\label{app:proof-dp}
\textbf{Algorithm.}
In the DP-SGD algorithm~\cite{abadi2016deep}, differential privacy is added to the training procedure by clipping a model's gradients below a certain threshold and adding noise to the sum using the following equations (slightly modified from their original presentation in Abadi et al.~\cite{abadi2016deep}):
\begin{align}
    \bar{g}_{_{i}} &= \frac{g_{_{i}}}{\max\Bigl\{1,\frac{\|g_{_{i}}\|_{2}}{C}\Bigl\}}\label{clipping-op} \\
    \tilde{g} &= \frac{1}{L} \left( \sum_{i}^{L} \bar{g}_{_{i}} + \mathcal{N}\left(0,\sigma^{2}C^{2}\mathbf{I}\right) \right) \label{noise-op}
\end{align}
where $g_{_{i}}$ is the gradient-vector of an arbitrary data point, C is the gradient norm bound, $\sigma$ is the noise scale, and L is the number of data points considered during a given step (i.e., batch size). By applying these operations and choosing $\sigma$ to be $\sqrt{2\log\frac{1.25}{\delta}} / \epsilon$, according to standard arguments~\cite{dwork2014algorithmic}, each step of DP-SGD is $(\epsilon,\delta)$-differentially private with respect to the batch. Moreover, for a training dataset of size $N$, sampling ratio equal to $\alpha = L / N$, and some overall number of steps $K$, using the moments account~\cite{abadi2016deep} the algorithm is $(O(\epsilon\alpha\sqrt{K}),\delta)$-differentially private with respect to the training dataset for an appropriate choice of noise scale and gradient norm bound. 

\textcolor{black}{Equation~\ref{noise-op} considers that each data point comes from the same dataset and has the same weight. 
In presence of multiple datasets (see~\ref{app:multiple_datasets}) we extend DP-SGD as follows
\begin{equation}\label{noise-op-weighted}
    \tilde{g} = \sum_{m=1}^{M} \frac{w_{m}}{L_{m}} \sum_{i=1}^{L_{m}} \bar{g}_{_{i,m}} + \mathcal{N}\left(0,\sigma^{2}C^{2}\mathbf{I} \right),
\end{equation}
where $L_{m}$ is the batch size for dataset $D_{m}$, $\bar{g}_{_{i,m}}$ is the gradient-vector of a data point in $L_{m}$ after the clipping operation \eqref{clipping-op} has been applied, and all other terms are the same as in~\eqref{noise-op}. In our analysis, we consider the case where $L_{m} = \alpha N_{m}$ and then $\alpha = \frac{L}{\sum_{i=1}^{m} N_{m}}$ is the sampling ratio.
}

\textit{
\textbf{Theorem~\ref{th:privacy-theorem}.} 
For some overall number of training steps, K, WERM minimized with DP-SGD (Eq.~\ref{noise-op-weighted}) is:
\begin{align*}
    &\Bigl( O(\epsilon_{_{T}}), \delta \Bigr)-\text{DP w.r.t. the training dataset} \: (D_{T}) \\
    &\Bigl( O(\epsilon_{_{R}}), \delta \Bigr)-\text{DP w.r.t. the reference dataset} \: (D_{R})
\end{align*}
where:
\begin{align*}
    \epsilon_{_{T}} &= \epsilon_{_{0}} \frac{1-w}{N_{T}} \\
    \epsilon_{_{R}} &= \epsilon_{_{0}} \frac{w}{N_{R}} \\
    0 &< \epsilon_{_{0}} < \min\Bigl(\frac{N_{T}}{1-w}, \frac{N_{R}}{w}\Bigl), \\
    \sigma &\geq \alpha \sqrt{K} \sqrt{2 \log \frac{1.25}{\delta}} \frac{C}{\epsilon_{_{0}}},
\end{align*}
$w$ is the reference data weight in~\eqref{w-erm-aggregated}, $N_{T}$ is the size of training data, $N_{R}$ is size of the reference data, K is the number of training steps, and $\sigma$, C, and $\alpha$ are the noise scale, gradient norm bound, and sampling ratio in DP-SGD, respectively. }

\begin{proof}
{\color{black}
We define $\tilde{g}_{_{m}} = \frac{w_{m}}{L_{m}} \sum_{i=1}^{L_{m}} \bar{g}_{_{i,m}}$ and start studying the case $\alpha = 1$. The $\ell_2$ sensitivity of $\tilde{g}_{_{m}}$ w.r.t.~a point in the dataset is $\Delta \tilde{g}_{_{m}} = 2 \frac{w_{m}}{N_{m}} C$.

Let $\epsilon_{_{1}} = \epsilon_{_{0}} \frac{w_{1}}{N_{1}}, \epsilon_{_{2}} = \epsilon_{_{0}} \frac{w_{2}}{N_{2}}, \dots, \epsilon_{_{M}} = \epsilon_{_{0}} \frac{w_{_{M}}}{N_{_{M}}}$ for some $\epsilon_{_{0}}>0$ such that $\max \{\epsilon_{_{1}}, \epsilon_{_{2}}, \dots, \epsilon_{_{M}}\} < 1$. We observe that with this choice $\frac{\Delta \tilde{g}_{_{1}}}{\epsilon_{_{1}}} = \frac{\Delta \tilde{g}_{_{2}}}{\epsilon_{_{2}}} =~\dots~= \frac{\Delta \tilde{g}_{_{M}}}{\epsilon_{_{M}}}=\frac{C}{\epsilon_0}$.
}


\textcolor{black}{
Reasoning as in the proof of Theorem 3.22 in Dwork and Roth~\cite{dwork2014algorithmic}, we conclude that if $\sigma \geq \sqrt{2 \log \frac{1.25}{\delta}} \frac{\Delta \tilde{g}_{_{m}}}{\epsilon_{_{m}}} = \sqrt{2 \log \frac{1.25}{\delta}} \frac{C}{\epsilon_{0}}$ then a step of the algorithm is:
\begin{align}
    &(\epsilon_{_{m}},\delta)-\text{DP w.r.t. the training dataset} \: D_{m}.
\end{align}
When $\alpha<1$, i.e., the batch size is smaller than the dataset size, we can invoke the privacy amplification theorem and each step of the algorithm becomes:
\begin{align}
    &\Bigl(O\Bigl(\epsilon_{_{m}}\alpha\Bigr),\delta\alpha\Bigr)-\text{DP w.r.t.} \: D_{m}.
\end{align}
Applying the moments account technique in~\cite{abadi2016deep} we see that over the whole training procedure for $K$ steps, the algorithm is:
\begin{align}
    &\Bigl( O\Bigl(\epsilon_{_{m}} \alpha \sqrt{K} \Bigr), \delta \Bigr)-\text{DP w.r.t.} \: D_{m}.
\end{align}    
}
\end{proof}

\subsection{Proof of Theorem~\ref{th:gen-bound}}\label{app:proof-gen-bound}
As a matter of notation, we write $L_{D}(f_\theta)$ as $L_{D}(\theta)$ to mean that the loss is evaluated over a model parameterized by $\theta$. 
{\color{black}
Also for this proof we consider the more general scenario with multiple datasets introduced in~\ref{app:multiple_datasets}. We will refer to additional auxiliary datasets $\hat{D}_{m} = \{\hat{z}^{(i)}_{m}, i \in [N_{m}]\}$ and  $\hat{D} = \bigcup_{m=1}^{M} \hat{D}_{m}$.
}

The proof of Theorem~\ref{th:gen-bound} relies on the following lemma, which we prove at the end of this section.

\textit{
\textbf{Lemma 1}
Under the assumption that the loss function is bounded, using our weighted ERM with training and reference data, it follows that:
\begin{equation*}
    \begin{aligned}
    \mathbb{E}_{D\sim \mathbb D^N} &\Bigl[\sup_{\theta \in \Theta} \Big|L_{\mathbb{D}}(\theta) - L^{w}_{D}(\theta) \Big|\Bigr] \\
    &\leq 2 \: \sqrt{\frac{\text{VCdim}(\Theta)}{N_{\text{eff}}}} \: \cdot \: \sqrt{\gamma_{2} + \log\Biggl(\frac{N}{\text{VCdim}(\Theta)}\Biggr)}
    \end{aligned}
\end{equation*}
where: 
\begin{align*}
L^{w}_{D}(\theta) &= (1-w) L_{D_{T}}(\theta) + w L_{D_{R}}(\theta), \\
\gamma_{2} &= \max \Bigl\{\frac{4}{\text{VCdim}(\Theta)}, 1 \Bigr\}, \\
N_{\text{eff}} &= \left[\frac{(1 - w)^{2}}{N_{T}} + \frac{w^{2}}{N_{R}} \right]^{-1},
\end{align*}
$D_{T},D_{R} \sim \mathbb{D}$, $N_{T} = |D_{T}|$, $N_{R} = |D_{R}|$, $D = D_{T} \cup D_{R}$, $N = |D|$, $\theta$ is a hypothesis in model class $\Theta$, and VCdim($\Theta$) is VCdim($\Theta$) is the VC-dimension of hypothesis class $F_{\Theta} = \{ f_{\theta} : \theta \in \Theta \} $, as defined in Shalev-Shwartz and Ben-David~\cite{shalev2014understanding}.}

\textit{
\textbf{Theorem~\ref{th:gen-bound}}
Under the assumption that the loss function is 
bounded in the range [0, 1], it follows that:
\begin{equation}\label{eq:gen-bound-formal}
    \begin{aligned}
    L_{\mathbb{D}}(\theta_{\text{WERM}}) \leq &\min_{\theta \in \Theta} L_{\mathbb{D}}(\theta) \\
    & + 2 \: \sqrt{\frac{\text{VCdim}(\Theta)}{N_{\text{eff}}}} \: \cdot \: \sqrt{\gamma_{2} + \log\Biggl(\frac{N}{\text{VCdim}(\Theta)}\Biggr)} \\
    & + \sqrt{\frac{2 \ln 2 / \delta}{N_{\text{eff}}}}
    \end{aligned}
\end{equation}
with probability $\geq 1 - \delta$,  where: 
\begin{align*}
\theta_{\text{WERM}} &= \argmin_{\theta \in \Theta} L^{w}_{D}({\theta}) \\
L^{w}_{D}(\theta) &= (1-w) L_{D_{T}}(\theta) + w L_{D_{R}}(\theta), \\
\gamma_{2} &= \max \Bigl\{\frac{4}{\text{VCdim}(\Theta)}, 1 \Bigr\}, \\
N_{\text{eff}} &= \left[\frac{(1 - w)^{2}}{N_{T}} + \frac{w^{2}}{N_{R}} \right]^{-1},
\end{align*}
$D_{T}\sim \mathbb D^{N_T}$, $D_{R} \sim \mathbb{D}^{N_R}$, 
$D = D_{T} \cup D_{R}$, $N = |D|$, 
and VCdim($\Theta$) is the VC-dimension of hypothesis class $F_{\Theta} = \{ f_{\theta} : \theta \in \Theta \} $.}

\begin{proof} 
We denote:
\begin{equation*}
    A(N_{\text{eff}}) = 2 \: \sqrt{\frac{\text{VCdim}(\Theta)}{N_{\text{eff}}}} \: \cdot \: \sqrt{\gamma_{2} + \log\Biggl(\frac{N}{\text{VCdim}(\Theta)}\Biggr)}
\end{equation*}
Lemma 1 proves the following:
\begin{equation*}
    \mathbb{E}_{D\sim \mathbb D^N} \Bigl[\sup_{\theta \in \Theta} \Big|L_{\mathbb{D}}(f_\theta) - L^{w}_{D}(f_\theta) \Big|\Bigr] \leq A(N_{\text{eff}})
\end{equation*}
In this proof, we will use the weaker bound:
\begin{equation}\label{eq:weaker-bound}
    \mathbb{E}_{D\sim \mathbb D^N} \Bigl[\sup_{\theta \in \Theta} \Big(L_{\mathbb{D}}(f_\theta) - L^{w}_{D}(f_\theta) \Big)\Bigr] \leq A(N_{\text{eff}}).
\end{equation}
We observe that changing a point in dataset $D_{m}$ leads $L_{\mathbb{D}}(\theta) - L^{w}_{D}(\theta)$ to change by at most $\frac{w_{m}}{N_{m}}$ in absolute value. In fact, call $\tilde{D}$ the dataset where a single point $z_{m}^{(i)}$ in $D$ has been replaced by point $\tilde{z}_{m}^{(i)}$:
\begin{align}
    \Big| &\sup_{\theta \in \Theta} L_{\mathbb{D}}(\theta) - L^{w}_{D}(\theta) - 
    \left( \sup_{\theta \in \Theta} L_{\mathbb{D}}(\theta^{\prime}) - L^{w}_{\tilde{D}}(\theta^{\prime}) \right)\Big| \\
    &\leq \Big|\sup_{\theta \in \Theta} \left( L_{\mathbb{D}}(\theta) - L^{w}_{D}(\theta) - L_{\mathbb{D}}(\theta) + L^{w}_{\tilde{D}}(\theta) \right) \Big| \\
    &= \Big| \sup_{\theta \in \Theta} \Big( L^{w}_{\tilde{D}}(\theta) - L^{w}_{D}(\theta) \Big) \Big| \\
    &= \Big| \sup_{\theta \in \Theta} \frac{w_{m}}{N_{m}} \Big( \ell(\theta, \tilde{z}^{(i)}_{m}) - \ell(\theta, z^{(i)}_{m} )\Big) \Big| \\
    &= \sup_{\theta \in \Theta} \Big| \frac{w_{m}}{N_{m}} \Big( \ell(\theta, \tilde{z}^{(i)}_{m}) - \ell(\theta, z^{(i)}_{m} )\Big) \Big| \\
    &\leq \frac{w_{m}}{N_{m}}
\end{align}
We can then apply McDiarmid’s inequality~\cite{shalev2014understanding} and we obtain:
\begin{align}
    \text{Prob} \Big( &\sup_{\theta \in \Theta} \Big( L_{\mathbb{D}}(\theta) - L^{w}_{D}(\theta) \Big) \nonumber\\
    &\le \mathbb{E} \Big[ \sup_{\theta \in \Theta} L_{\mathbb{D}}(\theta) - L^{w}_{D}(\theta) \Big] + \epsilon \Big) \nonumber\\
    &\ge 1 - \exp\Bigg( \frac{-2\epsilon^{2}}{\sum_{m=1}^{M} \sum_{i=1}^{N_{m}} \Big(\frac{w_{m}}{N_{m}}\Big)^{2}} \Bigg) \nonumber\\
    &= 1 - \exp\Big( -2 \epsilon^{2} N_{\text{eff}} \Big)
\end{align}
If we let $\delta = \exp\Big( -2 \epsilon^{2} N_{\text{eff}} \Big)$ we obtain:
\begin{align}
    \text{Prob} \Big( &\sup_{\theta \in \Theta} \Big( L_{\mathbb{D}}(\theta) - L^{w}_{D}(\theta) \Big) \nonumber \\
    &< \mathbb{E} \Big[ \sup_{\theta \in \Theta} L_{\mathbb{D}}(\theta) - L^{w}_{D}(\theta) \Big] \nonumber\\
    &+ \sqrt{\frac{1}{2 N_{\text{eff}}} \ln \frac{1}{\delta}} \Big) \geq 1 - \delta
\end{align}
and using~\eqref{eq:weaker-bound}:
\begin{align}
    \text{Prob} \Big( &\sup_{\theta \in \Theta} \Big( L_{\mathbb{D}}(\theta) - L^{w}_{D}(\theta) \Big) \leq \nonumber\\
    &A(N_{\text{eff}}) + \sqrt{\frac{1}{2 N_{\text{eff}}} \ln \frac{1}{\delta}} \Big) \geq 1 - \delta
\end{align}
As this is true for the $\sup_{\theta \in \Theta} L_{\mathbb{D}}(\theta) - L^{w}_{D}(\theta)$ it is true in particular for $\theta_{\text{WERM}}$:
\begin{equation}\label{eq:eq-2}
   \text{Prob}\left(   L_{\mathbb{D}}(\theta_{\text{WERM}}) - L^{w}_{D}(\theta_{\text{WERM}}) \leq A(N_{\text{eff}}) + \sqrt{\frac{1}{2 N_{\text{eff}}} \ln \frac{1}{\delta}} \right) \geq 1 - \delta.
\end{equation}

Consider $\theta^{\ast} \in \argmin_{\theta \in \Theta} L_{\mathbb{D}}(\theta)$:
\begin{equation*}
    L^{w}_{D}(\theta^{\ast}) = \sum_{m=1}^{M} \sum_{i=1}^{N_{m}} \frac{w_{m}}{N_{m}} \ell(\theta^{\ast}, z_{m}^{(i)}),
\end{equation*}
then it is a sum of independent random variables $\gamma_{m}^{(i)} = \frac{w_{m}}{N_{m}} \ell(\theta^{\ast}, z_{m}^{i}) \in \Big[0, \frac{w_{m}}{N_{m}} \Big]$.
By applying Hoeffding's inequality:
\begin{align}
    &\text{Prob}\Big( L^{w}_{D}(\theta^{\ast}) - L_{\mathbb{D}}({\theta^{\ast}}) \leq \epsilon \Big)\\
    &\geq 1 - \exp \Bigg( - \frac{2\epsilon^{2}}{\sum_{m=1}^{M}\sum_{i=1}^{N_{m}} \Big(\frac{w_{m}}{N_{m}} \Big)^{2}} \Bigg) \\
    &= 1 - \exp(-2 \epsilon^{2} N_{\text{eff}})
\end{align}
Similarly to above, we conclude:
\begin{equation}\label{eq:eq-3}
    L^{w}_{D}(\theta^{\ast}) - L_{\mathbb{D}}(\theta^{\ast}) \leq \sqrt{\frac{1}{2 N_{\text{eff}}} \ln \frac{1}{\delta}} \: \text{w.p.} \geq 1 - \delta 
\end{equation}
Both \eqref{eq:eq-2} and \eqref{eq:eq-3} hold w.p. $1 - 2\delta \: (P(A~\cap~B) = P(A) + P(B) - P(A~\cup~B) \geq P(A) + P(B) -1)$, thus:
\begin{align*}
    &L_{\mathbb{D}}(\theta_{\text{WERM}}) - L_{\mathbb{D}}(\theta^{\ast}) \\
    &= L_{\mathbb{D}}(\theta_{\text{WERM}}) - L^{w}_{D}(\theta_{\text{WERM}}) + L^{w}_{D}(\theta_{\text{WERM}}) \\
    & \phantom{space}- L^{w}_{D}(\theta^{\ast})  
    + L^{w}_{D}(\theta^{\ast}) - L_{\mathbb{D}}(\theta^{\ast}) \\
    &\leq L_{\mathbb{D}}(\theta_{\text{WERM}}) - L_{D}^{w}(\theta_{\text{WERM}}) + L^{w}_{D}(\theta^{\ast}) - L_{\mathbb{D}}(\theta^{\ast}) \\
    &\leq A(N_{\text{eff}}) + 2 \sqrt{\frac{1}{2 N_{\text{eff}}} \ln \frac{1}{\delta}} \\
    &\text{w.p.} \geq 1 - 2\delta
\end{align*}
Replacing $\delta$ by $\frac{\delta}{2}$, we obtain the theorem.
\end{proof}

\subsubsection{Proof of Lemma 1}

\begin{proof} 
Substituting the expectation over $L_{\hat{D}}(\theta)$ for $L_{\mathbb{D}}(\theta)$ and using Jensen's inequality, it follows that:
\begin{align}
     &\mathbb{E}_{D} \left[\sup_{\theta \in \Theta} \Big|L_{\mathbb{D}}(\theta) - L^{w}_{D}(\theta)\Big|\right] \\
     &\leq \mathbb{E}_{D,\hat{D}} \left[\sup_{\theta \in \Theta} \Big|L_{\hat{D}}(\theta) - L^{w}_{D}(\theta)\Big|\right] \\
     &= \mathbb{E}_{D,\hat{D}} \left[\sup_{\theta \in \Theta} \Big| \sum_{m=1}^{M} \sum_{i=1}^{N_{m}} w_{m} \left(\ell(\theta;z_{m}^{(i)}) - \ell(\theta;\hat{z}^{(i)}_{m}) \right) \Big| \right] \\
     &= \mathbb{E}_{D,\hat{D}} \mathbb{E}_{\sigma} \left[\sup_{\theta \in \Theta} \Big| \sum_{m=1}^{M} \sum_{i=1}^{N_{m}} \sigma_{m}^{(i)} \cdot w_{m} \left(\ell(\theta;z_{m}^{(i)}) - \ell(\theta;\hat{z}^{(i)}_{m}) \right) \Big| \right]
\end{align}
where $\sigma_{m}^{(i)}$ is a random variable drawn from the uniform distribution over $\{\pm 1\}$ that is uniquely sampled for each $m \in [M]$ and $i \in [N_{m}]$. Next, we fix $D$ and $\hat{D}$ and let C be the instances appearing in the two datasets. As defined in Definition 6.2 from Shalev-Shwartz and Ben-David~\cite{shalev2014understanding}, we assign $\Theta_{C}$ to be the restriction of $\Theta$ to C. Thus:
\begin{equation}
    \begin{aligned}
    \mathbb{E}_{D} &\left[\sup_{\theta \in \Theta} \Big|L_{\mathbb{D}}(\theta) - L^{w}_{D}(\theta)\Big|\right] \\
    &\leq \mathbb{E}_{D, \hat{D}} \mathbb{E}_{\sigma} \left[\sup_{\theta \in \Theta_{C}} \Big| \sum_{m=1}^{M} \sum_{i=1}^{N_{m}} \sigma_{m}^{(i)} \cdot w_{m} \left(\ell(\theta;z_{m}^{(i)}) - \ell(\theta;\hat{z}^{(i)}_{m}) \right) \Big| \right]
    \end{aligned}
\end{equation}
We fix some $\theta \in \Theta_{C}$ and denote $\gamma_{m}^{(i)} = \sigma_{m}^{(i)} \cdot w_{m} \left(\ell(\theta;z_{m}^{(i)}) - \ell(\theta;z^{\prime(i)}_{m}) \right)$ for $m \in [M]$ and $i \in [N_{m}]$. Without a loss of generality, by assuming the bound on the loss is between 0 and 1, we have that $\mathbb{E}[\gamma_{m}^{(i)}] = 0$ and $\gamma_{m}^{(i)} \in [-w_{m},w_{m}]$. Given that each $\gamma_{m}^{(i)}$ is an independent random variable, we invoke Hoeffding's inequality to say that:
\begin{equation}
    \begin{aligned}
    \mathbb{P} \Biggl[ \Bigg| \sum_{m=1}^{M} \sum_{i=1}^{N_{m}} \sigma_{m}^{(i)} \cdot w_{m} &\left(\ell(\theta;z_{m}^{(i)}) - \ell(\theta;\hat{z}^{(i)}_{m}) \right) \Bigg| \geq \rho \Biggr] \\
    &\leq 2\exp(-2N_{\text{eff}}\rho^2)
    \end{aligned}
\end{equation}
where $N_{\text{eff}} = \left( \sum_{m=1}^{M} \sum_{i=1}^{N_{m}} w_{m}^{2} \right)^{-1}$. Taking the the union bound over $\theta \in \Theta_{C}$, invoking Lemma A.4 in Shalev-Shwartz and Ben-David~\cite{shalev2014understanding}, we have:
\begin{equation}
    \begin{aligned}
    \mathbb{E} \Biggl[ \sup_{\theta \in \Theta_{C}} &\Big| \sum_{m=1}^{M} \sum_{i=1}^{N_{m}} \sigma_{m}^{(i)} \cdot w_{m} \left(\ell(\theta;z_{m}^{(i)}) - \ell(\theta;\hat{z}^{(i)}_{m}) \right) \Big| \Biggr] \\
    &\leq \frac{4 + \sqrt{\log(\Theta_{C})}}{\sqrt{2N_{\text{eff}}}} \\
    &\leq \frac{4 + \sqrt{\log(\tau_{\Theta}(N))}}{\sqrt{2N_{\text{eff}}}}
    \end{aligned}
\end{equation}
where $\tau_{\Theta}$ is the growth function of $\Theta$ as defined by Definition 6.9 in Shalev-Shwartz and Ben-David~\cite{shalev2014understanding}. Applying Lemma 6.10 in Shalev-Shwartz and Ben-David~\cite{shalev2014understanding} to upper bound $\tau_{\Theta}$ with respect to the VC-dimension and using the same steps as the proof of Lemma A.1 in Marfoq et al.~\cite{marfoq2022personalized}, we arrive at our generalization bound:
\begin{equation}
    \begin{aligned}
    \mathbb{E}_{D} &\Bigl[\sup_{\theta \in \Theta} \Big|L_{\mathbb{D}}(\theta) - L^{w}_{D}(\theta) \Big|\Bigr] \\
    &\leq 2 \: \sqrt{\frac{\text{VCdim}(\Theta)}{N_{\text{eff}}}} \: \cdot \: \sqrt{\gamma_{2} + \log\Biggl(\frac{N}{\text{VCdim}(\Theta)}\Biggr)}
    \end{aligned}
\end{equation}
where $\gamma_{2} = \max \Bigl\{\frac{4}{\text{VCdim}(\Theta)}, 1 \Bigr\}$. The result we present in the paper is for the case where $M=2$, which represents having access to training and reference data.

\end{proof}

\section{Implementation Details}
\subsection{Adversarial Regularization}\label{app:adv-reg}
The issue most relevant for our work is that the two players are not really alternating their optimization steps to solve the minmax problem, as described in the algorithm from Nasr et al.~\cite{nasr2018machine}. Instead, the code version updates the attack model for 52 batches, then the classifier for two batches, repeating this process 76 times, over the course of 20 rounds. By frequently alternating between the classifier and attack model, the training procedure in the code corresponds more closely to taking the entire gradient over the objective function and updating both models simultaneously. After evaluating different variants that utilize some aspects from the general description and other aspects from the code implementation, the best results are obtained by updating each model with a higher frequency.

Additionally, the released code contains a bug where not all batches are observed an equal number of times. To address this issue and remain faithful to the goal of alternating model updates with a higher frequency, we use the following procedure:
\begin{enumerate}
    \item Define the ratio that the attack model should be trained compared to the classifier
    \item For each batch in an epoch, sample from a Bernoulli distribution using the defined training ratio
    \item If a 0 is drawn, train the attack model; if a 1 is drawn, train the classifier
    \item Repeat steps 1-3 until the conclusion of the epoch
\end{enumerate}
Taking into account the bug, we measured the effective attack model to classifier training ratio to be 22:1. We produce the same results using the above training procedure with a ratio of 20:1.

As a final point, in the original formulation, the attack model inputs the raw feature vectors during training. Alternatively, in the released code version, the attack model only has access to the ground-truth labels and confidence-vector outputs. We follow the code implementation in our experiments.

\subsection{MMD-based Regularization}\label{app:mmd}
As there is no released code associated with MMD~\cite{li2021membership}, we implement the method ourselves based on the paper's description. It is written that ``In the implementation of our MMD loss, we reduce the difference between the probability vector distributions of members and non-members for the same class. That is, a batch of training samples and a batch of validation samples in the same class are used together to compute
the MMD score.'' Typically, during training one calculates the gradients of the objective function with respect to a mini-batch that is sampled uniformly from the training data. Selecting a mini-batch the only has data with the same label results in a model training procedure that stagnates and fails to learn. How does MMD ensure that the regularization term only compares data with the same label? We reached out to the authors to ask for their exact training strategy but did not receive a reply. Accordingly, we propose the following procedure that is faithful to the description provided in the text:
\begin{enumerate}
    \item Sample a mini-batch uniformly from the training data using a sufficiently large batch size to include multiple instances of each label
    \item Sample an equally sized mini-batch uniforming of reference data
    \item Identify all unique labels in the training data mini-batch
    \item For each distinct label in the training data mini-batch, calculate the MMD between instances with this label in the training and reference mini-batches
    \item Use the average MMD across all distinct labels as the regularization term
\end{enumerate}
Additionally, as the value of the variance for the Gaussian kernel used in~\eqref{eq:mmd} is not specified, we experimented with several values before deciding to use 1.0.

\section{Dataset Descriptions}\label{dataset-desc}
\subsection{Purchase100}
This dataset, derived from the ``acquire valued shopper'' challenge on Kaggle, consists of shopping records for several thousand individuals.~\footnote{https://www.kaggle.com/competitions/acquire-valued-shoppers-challenge/overview} Using this dataset, participants aim to find discounts that can attract shoppers to buy new products. We use the same pre-processed version of the dataset as in Nasr et al.~\cite{nasr2018machine}. In total, there are 197,324 data points, where each entry contains 600 binary features that indicate whether the shopper has purchased a certain item. Based on these binary features vectors, the data is clustered into 100 classes that represent distinct categories of shoppers. The prediction task is to determine the class associated to each shopper.

\subsection{Texas100}
This dataset, released by the Texas Department of State Health Services, consists of information regarding inpatient stays at several health facilities. Each record encodes the external cause of injury (e.g., suicide, drug misuse), the diagnosis (e.g., schizophrenia, depression), the procedures underwent by the patient (e.g., X-ray, surgery), the length of stay, personal information relating to the patient (e.g., gender, age, race), and hospital-specific identifiers (e.g., hospital ID). We use the same pre-processed version of the dataset as in Nasr et al.~\cite{nasr2018machine}. In total, there are 67,330 data points, where each entry contains 6,170 binary features that indicate whether a patient underwent any of the 100 most common medical procedures. Based on these binary features vectors, the data is clustered into 100 classes that represent distinct categories of patients. The prediction task is to determine the class associated to each patient.

\subsection{CIFAR100}
This dataset is a major benchmark generally used to in image recognition. It contains 60,000 images and 100 classes. Each image is composed of 32 x 32 color pixels.

\section{Membership Inference Attack Details}\label{app:mi-attacks}
\subsection{Assumptions in Membership Inference Attacks}\label{app:mia-assumptions}
\textcolor{black}{In Table~\ref{table:attacks}, we present a list of the most well-known MIAs categorized by their assumption settings.}

\begin{table*}[hbt!]
\begin{tabular}{ |p{4.5cm}|p{3cm}|p{2.5cm}|p{3.3cm}|p{2.5cm}|}
\hline
attack name & ground-truth training data & population data & vector of attack & ground-truth label \\
\hline
Gap Attack~\cite{yeom2018privacy} & no & no & predicted label & yes \\
Confidence Attack \cite{yeom2018privacy} & no & no & largest confidence value & no \\
Entropy Attack \cite{song2021systematic,shokri2017membership} & no &  no & confidence-vector & no \\
Modified-entropy Attack \cite{song2021systematic} & no & no & confidence-vector & yes \\ 
Neural Network Attack \cite{shokri2017membership} & yes & yes & confidence-vector & yes \\
Distillation Attack~\cite{ye2021enhanced} & no & yes & model loss & yes \\
Likelihood Ratio Attack (LiRA)~\cite{carlini2022membership} & maybe & yes & model loss & yes \\
Leave-one-out Attack~\cite{ye2021enhanced} & yes & yes & model loss & yes \\
\hline
\end{tabular}
\caption{Analyzing black-box membership inference attacks by assumptions. The term ``maybe'' means that using additional knowledge derived from this assumption can be directly integrated into the attack to improve performance.} \label{table:attacks}
\end{table*}

\subsection{Design Details}\label{app:mia-design}
\subsubsection{Gap Attack}
In the standard case where an attack is evaluated using two equally sized datasets, $D_{T}^{\text{adv}}$ and $D_{\overline{T}}^{\text{adv}}$, the overall attack accuracy for he gap attack~\eqref{gap-attack-single} can be computed as follows:
\begin{equation}\label{gap-attack}
    \text{acc}_\text{gap attack} = \frac{1}{2} + \frac{\text{acc}_{T} - \text{acc}_{\overline{T}}}{2}
\end{equation}
where $\text{acc}_{T}$ and $\text{acc}_{\overline{T}}$ correspond to the target model's prediction accuracy computed over $D_{T}^{\text{adv}}$ and $D_{\overline{T}}^{\text{adv}}$, respectively~\cite{choquette2021label}. As a model's loss is negatively correlated with its accuracy, a model that has poor generalization (i.e., relatively low train loss compared to test loss) will be be vulnerable to the gap attack. Thanks to its simplicity and demonstrated effectiveness~\cite{song2021systematic}, the gap attack has been suggested as a baseline attack against which all proposed defenses should be evaluated~\cite{choquette2021label}.

\subsubsection{Threshold-based Attacks}
When the adversary has access to a model's entire confidence-vector output, various other metrics, such as the entropy of the output distribution, can be utilized instead of the confidence value for threshold-based attacks~\cite{shokri2017membership}.
Estimating a class-dependent threshold or class-specific metric (e.g., modified-entropy~\cite{song2021systematic}) is possible when the adversary knows the target data points' ground-truth labels. It is sometimes assumed that the adversary has access to additional data from the underlying distribution in the form of known member and non-member data, which allows for the estimation of more precise thresholds~\cite{yeom2018privacy,shokri2017membership,song2021systematic}.

Threshold-based attacks exploit the phenomenon that 
target model's outputs for training data are usually distinguishable from target model's outputs for non-training data (e.g., confidence values can be more skewed in the first case).
Moreover, all variations of threshold-based attacks, as well as other types of MIAs, can be improved by querying the target model with transformed or augmented versions of target data points and predicting membership based on the aggregated output~\cite{choquette2021label, carlini2022membership}.

\section{Additional Figures/Tables}\label{app:additional-figs}
\begin{figure*}
  \centering
  \includegraphics[width=1\linewidth]{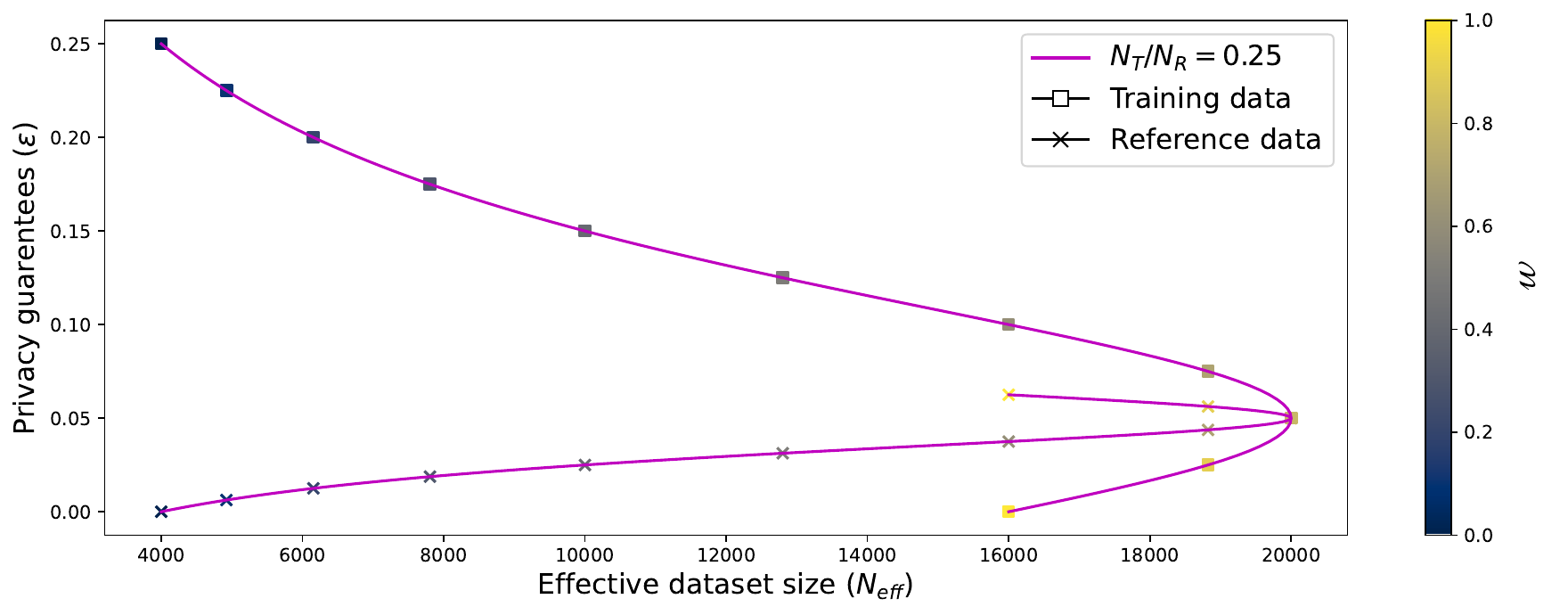}
  \includegraphics[width=1\linewidth]{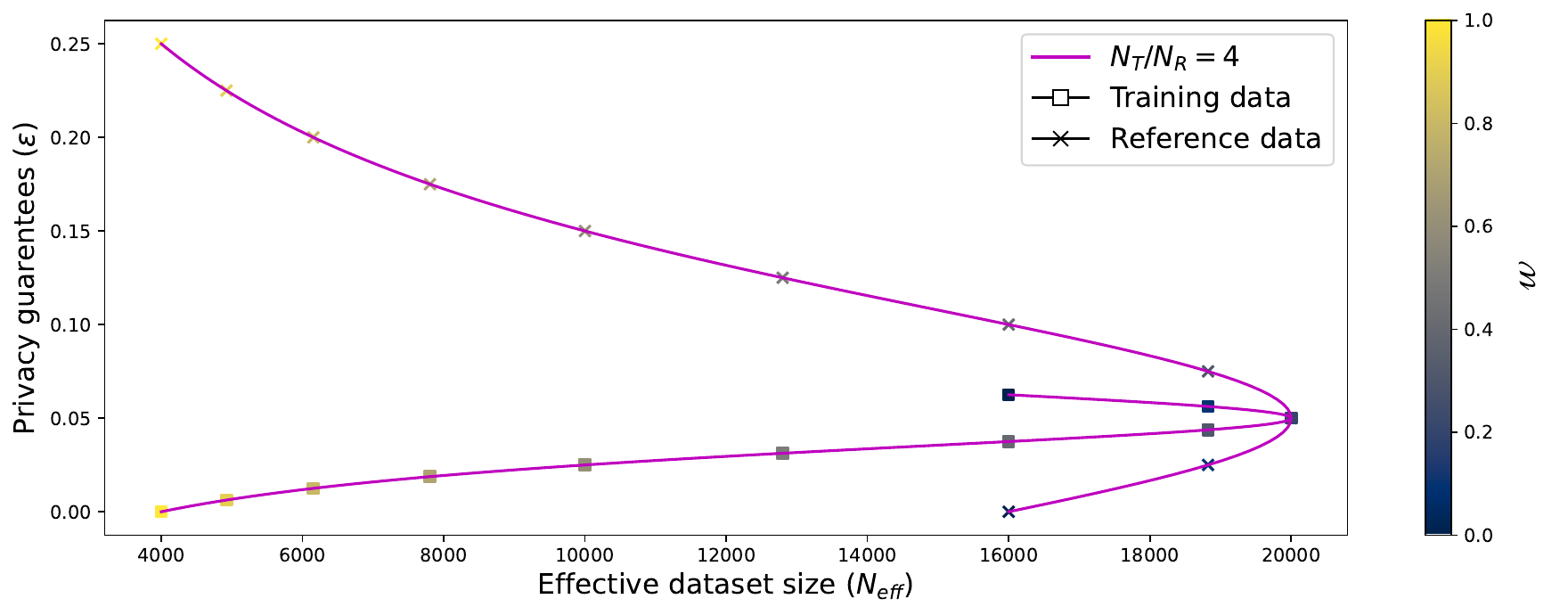}
  \caption{\textcolor{black}{Theoretical bounds for $\frac{N_T}{N_R} = 0.25$ and its inverse $\frac{N_T}{N_R} = 4$.}}
  \label{fig:inverse-theoretical-curves}
\end{figure*}

\begin{figure*}
     \centering
     \includegraphics[width=1\linewidth]{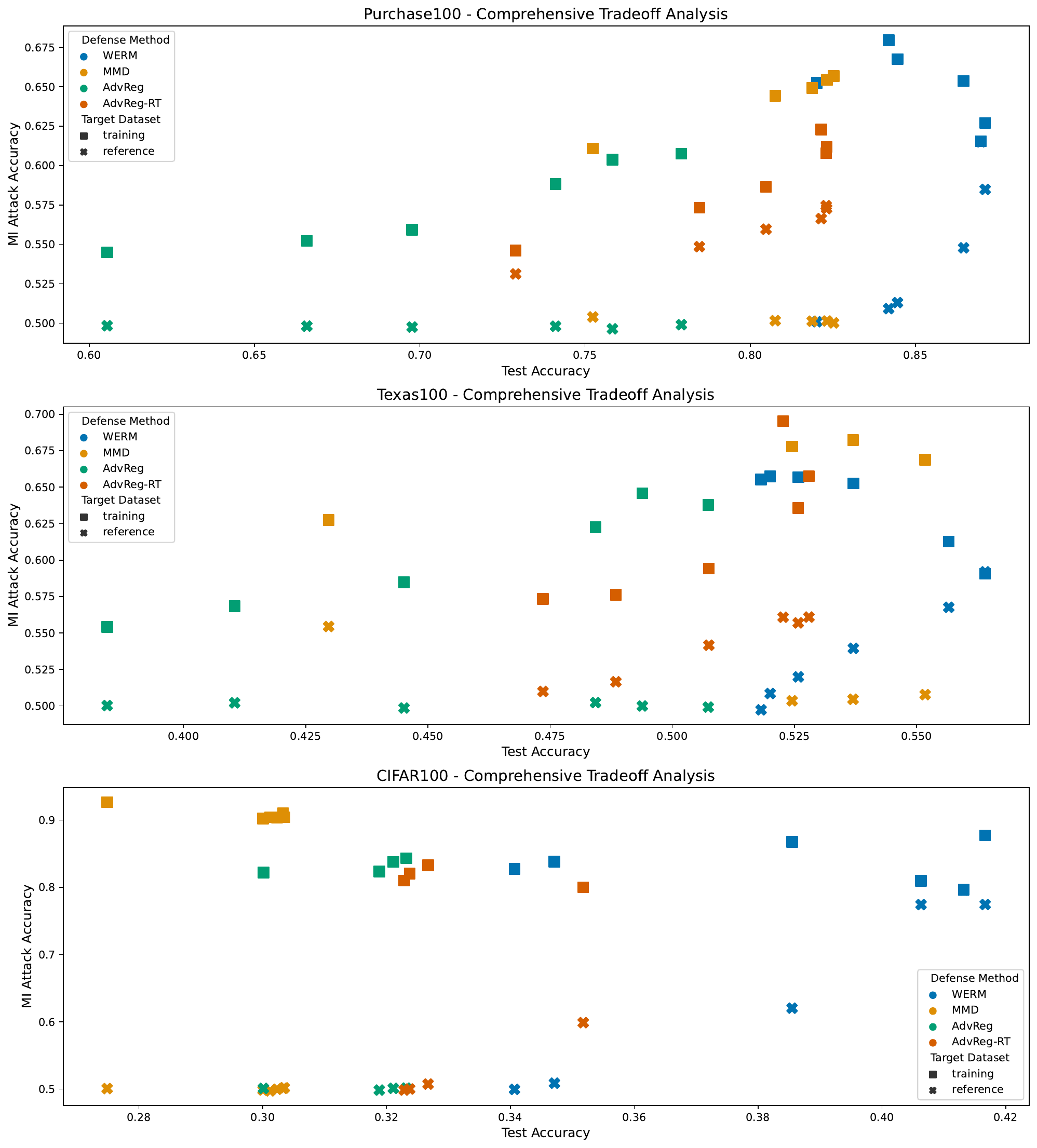}
    \caption{Utility-Privacy tradeoffs obtained by various empirical privacy defenses for the Purchase100, Texas100, and CIFAR100 datasets. The test accuracy of a defended classifier is measured using unseen test data and the MIA accuracy on training and reference data is evaluated with a threshold-based using confidence values (Eq.~\ref{thresh-attack}). Each point on the curve represents the evaluation of a model instance using a distinct regularization value (for AdvReg, AdvReg-RT, and MMD) and reference data weight value (for WERM). In these curves, the model instance is selected on the basis of validation data.}\label{fig:three-curves-valid}
    \Description{The utility-privacy curves are plotted for all of the empirical privacy defenses that we evaluate, where the model instance during training is selected on the basis of validation data.}
\end{figure*}

\begin{figure*}
     \centering
     \includegraphics[width=1\linewidth]{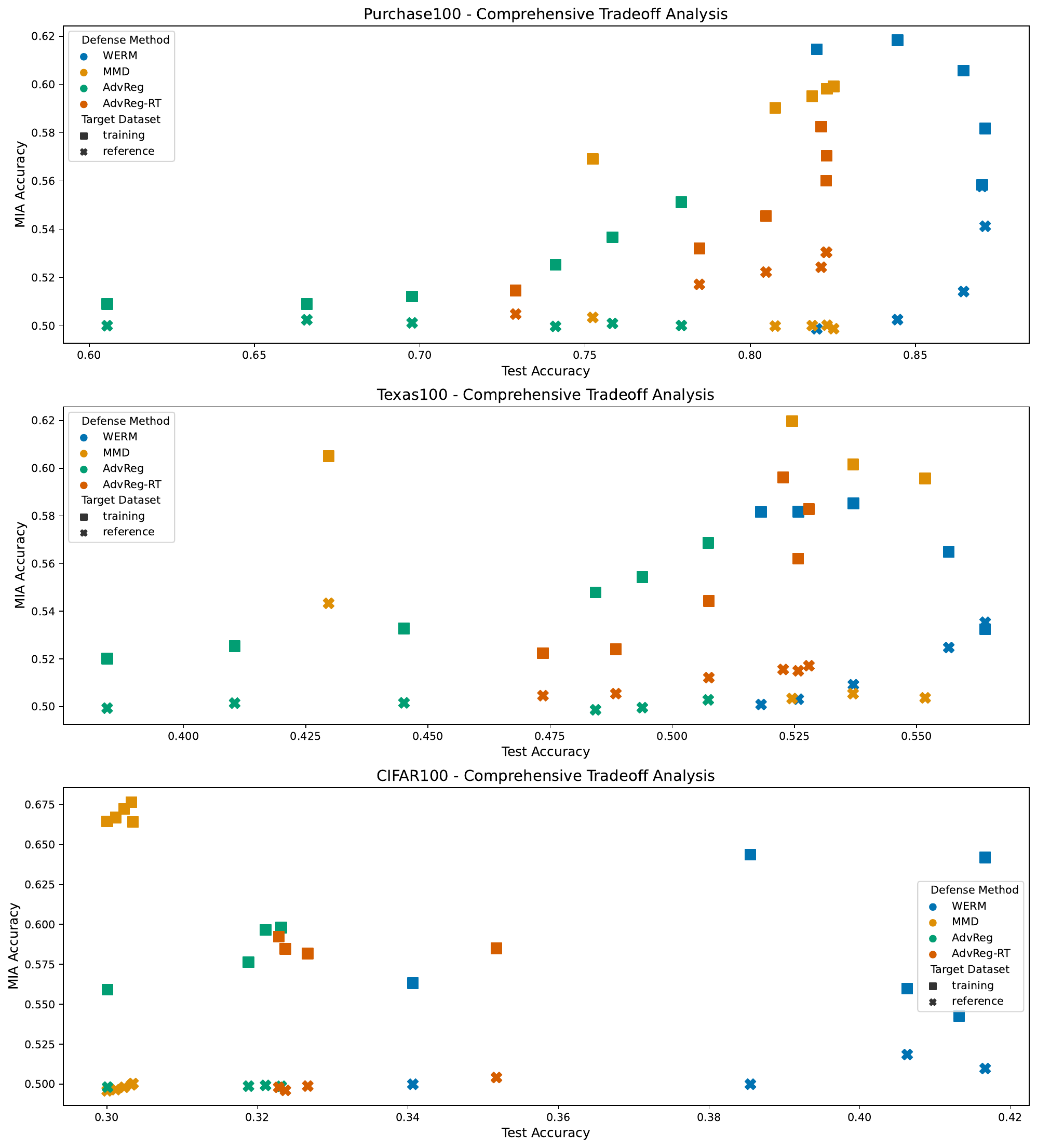}
    \caption{\textcolor{black}{Utility-Privacy tradeoffs obtained by various empirical privacy defenses for the Purchase100, Texas100, and CIFAR100 datasets. The test accuracy of a defended classifier is measured using unseen test data and the MIA accuracy on training and reference data is evaluated with a neural network attack~\cite{nasr2018machine,shokri2017membership}. Each point on the curve represents the evaluation of a model instance using a distinct regularization value (for AdvReg, AdvReg-RT, and MMD) and reference data weight value (for WERM). In these curves, the model instance is selected on the basis of validation data.}}\label{fig:three-curves-nn}
    \Description{The utility-privacy curves are plotted for all of the empirical privacy defenses that we evaluate, where the model instance during training is selected on the basis of validation data and MIA accuracy is measured using a neural network attack.}
\end{figure*}

\begin{figure*}
     \centering
     \includegraphics[width=1\linewidth]{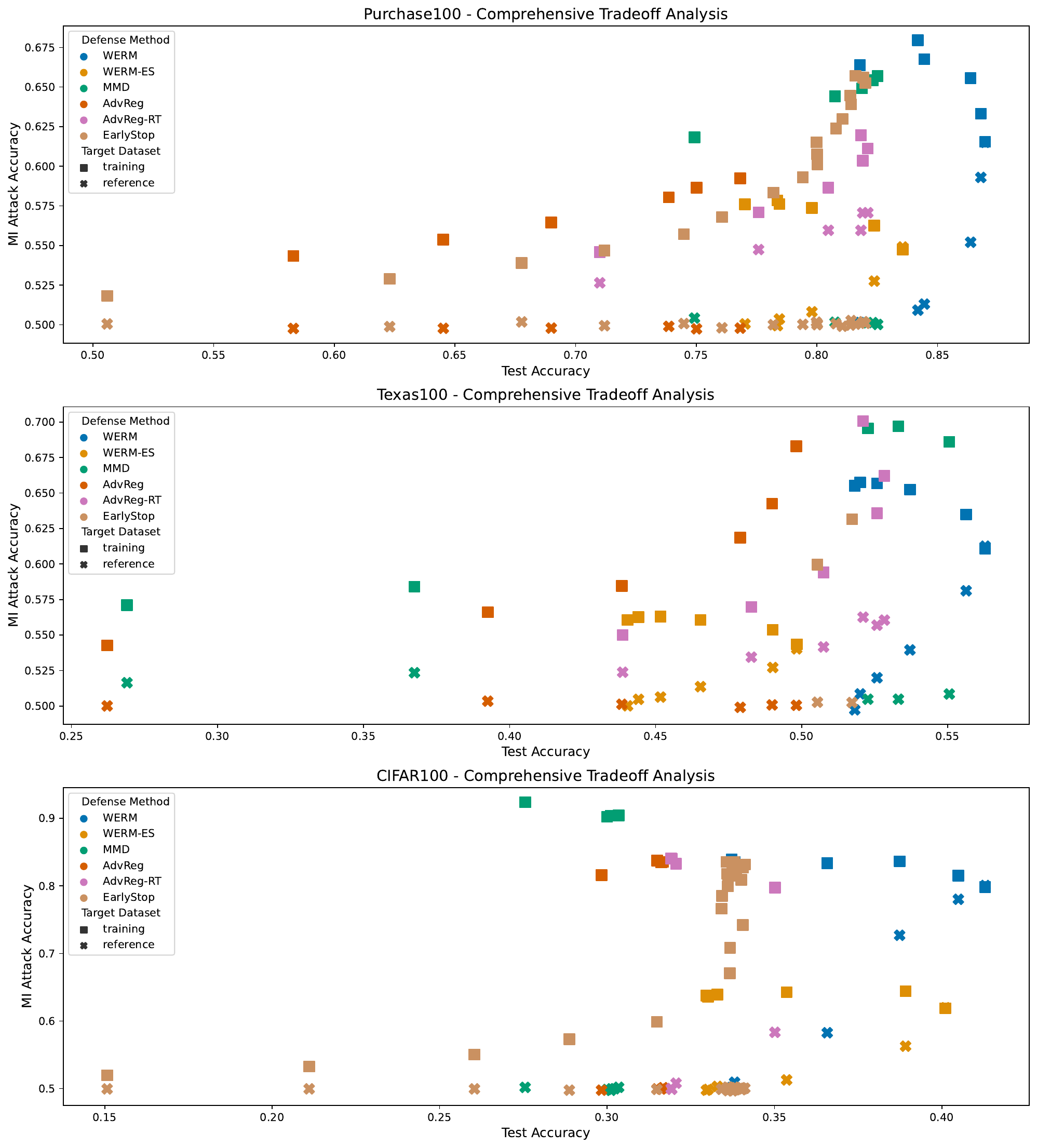}
    \caption{This figure examines the utility-privacy curves produced by all versions of empirical privacy defenses for the Purchase100, Texas100, and CIFAR100 datasets. The test accuracy of a defended classifier is measured using unseen test data and the MIA accuracy on training and reference data is evaluated with a threshold-based using confidence values (Eq.~\ref{thresh-attack}). Each point on the curve represents the evaluation of a model instance using a distinct regularization value.}\label{fig:three-curves-full}
    \Description{The utility-privacy curves are plotted for all of the empirical privacy defenses that we evaluate and compared against our WERM and WERM-ES baseline. The takeaway is that AdvReg and MMD cannot produce model instances that tradeoff reference data privacy for model utility or training data privacy, whereas our baseline is able to make these tradeoffs.}
\end{figure*}

\textcolor{black}{In Figure~\ref{fig:inverse-theoretical-curves}, we show the behavior for ${N_{T}}/{N_{R}} = 0.25$ and its reciprocal ${N_{T}}/{N_{R}} = 4$.} In Figure~\ref{fig:three-curves-valid}, we present the utility-privacy curves where the model instances are selected on the basis of validation data. \textcolor{black}{In Figure~\ref{fig:three-curves-nn}, we present the utility-privacy curves using a neural network attack to evaluate the MIA accuracy. We follow the same methodology as~\cite{nasr2018machine} in designing the neural network attack. The results do not seem better than the threshold-based MIAs we evaluate against. However, one must consider that our threshold-based MIAs optimize the threshold with the exact same information that the neural network attack learns from.} In Figure~\ref{fig:three-curves-full}, we present the utility-privacy curves including the curve for WERM-ES, without highlighting trends, and without removing low test accuracy outliers for AdvReg and MMD. \textcolor{black}{In Table~\ref{table:alignment-analysis-per-dataset}, we show a comparison of the Pearson correlation coefficient calculated for each dataset individually. We can see that WERM outperforms both AdvReg and MMD on all three datasets. The correlation difference is particularly significant on CIFAR100.}

\begin{table}
\caption{\textcolor{black}{Comparison of the Pearson Correlation Coefficient (PCC) between the training-reference data desired privacy ratio (as determined by the choice of $w$ or $\lambda$) and the empirical privacy ratio (as measured by a MIA) for WERM, AdvReg, and MMD. The coefficient is computed for the Purchase100, Texas100, and CIFAR100 datasets individually and across all datasets (i.e., the theoretical and empirical relative privacy values are aggregated before calculating the PCC).}}\label{table:alignment-analysis-per-dataset}
\centering
\begin{tabular}{c c c}
    \toprule
defense & dataset & Pearson Correlation Coefficient \\
    \midrule
\multirow{4}{*}{WERM}  
        & Purchase100   & 1.0       \\
        & Texas100      & 0.99      \\
        & CIFAR100      & 0.99      \\
        & Overall       & 0.84      \\    
    \addlinespace
    \hline
    \addlinespace
\multirow{4}{*}{AdvReg}
        & Purchase100   & 0.87      \\
        & Texas100      & 0.09      \\
        & CIFAR100      & -0.97     \\
        & Overall       & 0.07      \\     
    \bottomrule
    \addlinespace
\multirow{4}{*}{MMD}
        & Purchase100   & 0.93      \\
        & Texas100      & 0.86      \\
        & CIFAR100      & 0.49      \\
        & Overall       & 0.48      \\    
    \bottomrule
    \addlinespace
\end{tabular}
\end{table}

\end{document}